\documentclass[twocolumn,twocolappendix]{aastex631}

\usepackage[]{txfonts}
\usepackage{graphicx}
\usepackage{bm}
\usepackage{ascmac}

\graphicspath{{./Kondo/}}

\newcommand{\eqref}[1]{Equation~(\ref{#1})}
\newcommand{\figref}[1]{Figure~\ref{#1}}
\newcommand{\dfrac}[2]{ {\displaystyle\frac{#1}{#2}} }
\newcommand{\dint}{ {\displaystyle\int} }

\newcommand{\pr}[1]{\ensuremath{\left(#1\right)} }
\newcommand{\pd}[2]{\ensuremath{\dfrac{\partial #1}{\partial #2} }}
\newcommand{\pf}[2]{\ensuremath{\left( \dfrac{#1}{#2} \right)} }

\newcommand{\Alfven}{Alfv\'en }

\newcommand{\Am}{{\rm Am} }

\newcommand{\z}{$z$}
\renewcommand{\d}{{\rm d}}
\newcommand{\da}{{\rm d}a}

\newcommand{\dz}{\,{\rm d}z}

\newcommand{\cs}{c_{\rm s}}

\newcommand{\alppz}{\overline{\alpha_{z \phi}}}

\newcommand{\heat}{\rm heat}
\newcommand{\amax}{a_{\rm max}}
\newcommand{\amin}{a_{\rm min}}
\newcommand{\rhoint}{\rho_{\rm int}}

\newcommand{\um}{\ensuremath{\rm \,\mu m }}
\newcommand{\cm}{\ensuremath{\rm \,cm }}
\newcommand{\K}{\ensuremath{{\rm \,K }} }
\newcommand{\au}{\ensuremath{{\rm \,au }}}
\newcommand{\yr}{\ensuremath{\rm \,yr }}
\newcommand{\Msun}{\ensuremath{\rm \,M_{\sun}}}

\newcommand{\Mpyr}{ \ensuremath{\rm \,\Msun \,\yr^{-1} }}

\newcommand{\gcmcmcm}{\ensuremath{\rm \,g \, cm^{-3} }}
\newcommand{\ergs}{\ensuremath{\rm \,erg \, s^{-1} }}
\newcommand{\Myr}{ \ensuremath{\rm \,Myr}}

\definecolor{darkorange}{rgb}{1.0, 0.55, 0.0}

\begin{document}
\title{The Roles of Dust Growth in the Temperature Evolution and Snow Line Migration in Magnetically  Accreting Protoplanetary Disks
}

\shortauthors{Kondo et al.}

\def\myemail{kondo.k.al@m.titech.ac.jp}
\def\titech{Department of Earth and Planetary Sciences, Tokyo Institute of Technology, Meguro-ku, Tokyo 152-8551, Japan}
\def\tohoku{Astronomical Institute, Tohoku University, 6-3 Aramaki, Aoba-ku, Sendai 980-8578, Japan}

\author[0000-0002-1219-6499]{Katsushi Kondo}       \affiliation{\titech}   
\author[0000-0002-1886-0880]{Satoshi Okuzumi}  \affiliation{\titech}
\author[0000-0002-7002-939X]{Shoji Mori}    \affiliation{\tohoku}


\begin{abstract}
The temperature structure of protoplanetary disks provides an important constraint on where in the disks rocky planets like our own form. Recent nonideal magnetohydrodynamical (MHD) simulations have shown that the internal Joule heating associated with magnetically driven disk accretion is inefficient at heating the disk midplane. A disk temperature model based on the MHD simulations predicts that in a disk around a solar-mass young star, the water snow line can move inside the current Earth's orbit within 1 Myr after disk formation. However, the efficiency of the internal Joule heating depends on the disk's ionization and opacity structures, both of which are governed by dust grains. In this study, we investigate these effects by combing the previous temperature model for magnetically accreting disks with a parameterized model for the grain size and vertical distribution. Grain growth enhances the gas ionization fraction and thereby allows Joule heating to occur closer to the midplane. However, growth beyond 10 $\um$ causes a decrease in the disk opacity, leading to a lower midplane temperature. The combination of these two effects results in the midplane temperature being maximized when the grain size is in the range 10--100 $\um$. Grain growth to millimeter sizes can also delay the snow line's migration to the 1 au orbit by up to a few Myr. We conclude that accounting for dust growth is essential for accurately modeling the snow line evolution and terrestrial planet formation in magnetically accreting protoplanetary disks.
\end{abstract}%



\section{Introduction}\label{sec:intro}

The terrestrial planets in our solar system are thought to be significantly depleted in water compared to outer solar system bodies. 
The mass of the present Earth's ocean comprises only 0.02\% of the Earth's total mass \citep[e.g.,][]{C&S2010}, and estimates show that the present Earth would only contain water of at most 10 ocean masses even if the bulk mantle's water is taken into account \citep[e.g.,][]{Marty12}. The Earth's initial water content may be higher but is unlikely to well exceed 1 wt\%  \citep{Abe2000,Tagawa21}. The water contents of the other terrestrial planets, namely  Mercury, Venus, and Mars, are also suggested to be less than $1$ wt\% \citep[e.g.,][]{Lawrence2013,ET2007,Kurokawa2014}. In contrast, Neptune or comets, which would have formed in the outer part of the solar system, contain more than $10$ wt\% water \citep[e.g.,][]{Guillot2005,A'Hearn2011,Rotundi2015}. 

To constrain when, where, and how the terrestrial planets formed in the solar nebula, it is important to understand how the nebula's temperature structure evolved with time. In protoplanetary disks, water ice is stable outside the snow line, where the gas temperature equals the water sublimation temperature ($\sim 160$--$170\K$; \citealt{Hayashi1981}).
The relatively high water-to-rock ratios of comets suggest that the dust outside the solar nebula's snow line may have contained a large amount of water ice. Planetesimals may form early outside the snow line lost water after radiogenic heating \citep{Lichtenberg19}; however, they  may have eventually become icy planetary embryos by capturing icy pebble-sized particles transported from the nebula's outer region \citep{Sato2016}. 
Therefore, if the solar nebula's snow line existed well inside 1 au, one must assume that either the terrestrial planets or their building blocks formed in the nebula's innermost region and then migrated outward \citep{Ogihara15,Ogihara18};  they  formed after the formation of giant planets, which may have blocked inward drifting icy pebbles \citep{Morbidelli16}; or they grew into their current sizes through pebble accretion, in which the accreted pebbles may have lost ice in the heated planetary atmosphere \citep{Johansen21}.

Previous studies on the snow line evolution \citep[e.g.,][]{S&L2000,G&L2007,Oka2011,Z&J2015,BitschRaymond19} adopted the classical viscous accretion disk model with vertically uniform viscosity. This model effectively assumes that the viscous heating dominantly occurs around the disk midplane, and predicts that the viscous heating dominates over stellar irradiation heating at warms the midplane of the inner disk region with a large optical depth. In this model, viscous heating determines the evolution of the snow line location. 

However, the validity of using such a simple viscous accretion model to study the snow line evolution has been unclear because no known hydrodynamical instability is likely to produce strong turbulence in the inner protoplanetary disk region. It has been long recognized that the magnetorotational instability \citep[MRI;][]{B&H1991}, which is a magnetohydrodynamical instability generating strong turbulence in ionized Keplerian disks, is unlikely to operate in the  midplane of the $\ga 1~\rm au$ region because of strong Ohmic diffusion \citep{Gammie1996}. More recent studies \citep[e.g.,][]{B&S2013b,Lesur14,Gressel2015} have shown that ambipolar diffusion can also suppress the MRI turbulence around the disk surface, suggesting that the inner few au region would be entirely laminar unless other hydrodynamical instabilities produce turbulence. The Hall effect can amplify and suppress magnetic fields depending on the direction of the magnetic field threading the disk \citep[e.g.,][]{Sano02a,Sano02b,Kunz08,Bai2014,Lesur14,Bai2017}, but does not directly contribute to internal heating because of its non-dissipative nature \citep{Mori2019}. 
Instead, these studies propose that the disk accretion is driven by  magnetic disk winds that take away the disk's angular momentum \citep{B&S2013a}. 
In this paper, we call such disks magnetically accreting disks or more simply MHD disks.
The possibility remains that  non-MHD instabilities, in particular the convective overstability \citep[COV;][]{K&H2014} and zombie vortex instability \citep[ZVI;][]{Marcus2015}, produce a certain level of turbulence around the snow line \citep{Malygin2017,P&K2019,L&U2019}. However, simulations suggest that the midplane turbulence produced by these mechanisms would be weaker than full-fledged MRI turbulence \citep{Barranco2018,Raettig2021}. 

In magnetically accreting disks mentioned above, the internal heating process differs significantly from the viscous heating in the classical accretion disk model. First of all, it is the Joule heating, not the dissipation of turbulence, that dominates the heat generation in MHD accretion disks with a low ionization fraction \citep{Sano01}. \citet{Hirose11} already showed using MHD simulations including Ohmic diffusion that the disk's Joule heating region is localized to a thin current layer well above the midplane, where the ionization fraction is moderately high. 
Recently, \cite{Mori2019} have confirmed the finding by \citet{Hirose11} with MHD simulations including Ohmic, ambipolar, and Hall effects. Because the heat is generated at low optical depths,  accretion heating in magnetically accreting disks is much less efficient than in  classical viscous disks with vertically uniform viscosity.

Based on the simulation results of \citet{Mori2019}, \cite{Mori2021} constructed an empirical model that predicts the midplane temperature of a magnetically accreting protoplanetary disk from the disk's ionization structure and opacity. For disks around solar-type stars, they found that the snow line migrates inside  $1\au$ within the first $\sim 10^{5}~\rm yr$ after star formation. The classical viscous model also predicts inward migration of the snow line, but the snow line's arrival time at 1 au in the MHD disk model is an order of magnitude longer in the model assuming vertically uniform viscosity (see Figure 6 of \citealt{Mori2021}). The MHD model suggests that planetary embryos at $1\au$ could have accreted a larger amount of water ice \citep[see also][]{Sato2016}, strongly constraining scenarios for the inner solar system formation. 

However, \cite{Mori2021} employed a simplified dust model that could affect their conclusion. In their model, dust grains play two important roles. First, small grains are efficient at capturing ionized electrons and ions in the gas, and therefore the gas ionization fraction depends strongly on the abundance of such grains \citep[e.g.,][]{Sano01,Wardle2007,Okuzumi2009}. Second, {dust grains are the dominant source of disk opacity}, and their size and spatial distribution control the disk's cooling rate \citep[e.g.,][]{Oka2011}. 
\citet{Mori2021} assumed that submicron-sized grains are abundant in the disks\footnote{Specifically, the fiducial model of \citet{Mori2021} assumes $0.1\um$-sized grains with a dust-to-gas mass ratio of 0.01.}; 
however, it is more likely that most of the small grains in the inner few au region grow to larger solid particles or bodies in the first 1~Myr of disk evolution \citep[e.g.,][]{Birnstiel2010}. 
As the abundance of small dust grains decreases, the ionization fraction increases, whereas the disk opacity decreases.
Their net effect is not obvious, because an increased ionization fraction results in a Joule heating layer closer to the midplane, while a decreased opacity results in more efficient radiative cooling. 
\cite{Mori2021} were unable to study the net effect of the two in a self-consistent manner because they determined the disk's ionization faction and opacity independently.

In this paper, we investigate the effects of dust size distribution on the temperature and snow line evolution in magnetically accreting disks. 
We extend the MHD disk temperature model of \cite{Mori2021} so that one can compute the disk's ionization and opacity structure from a consistent dust size and vertical distribution. With this model, we study in detail whether dust growth can delay the early snow line migration found in the previous calculations by \citet{Mori2021}.

This paper is organized as follows. In Section \ref{sec:model}, we describe the model used in this study. In Section~\ref{sec:result}, we show the temperature structure and the snow line's arrival time at $1\au$ in MHD disks around a solar-mass star. In Section \ref{sec:dis}, we discuss the validity of our model and implications for terrestrial planet formation. A summary is given in Section~\ref{sec:summary}.

\section{Model}\label{sec:model}
\begin{figure*}
    \centering
    \includegraphics[bb=0 0 1917 647,width=\hsize]{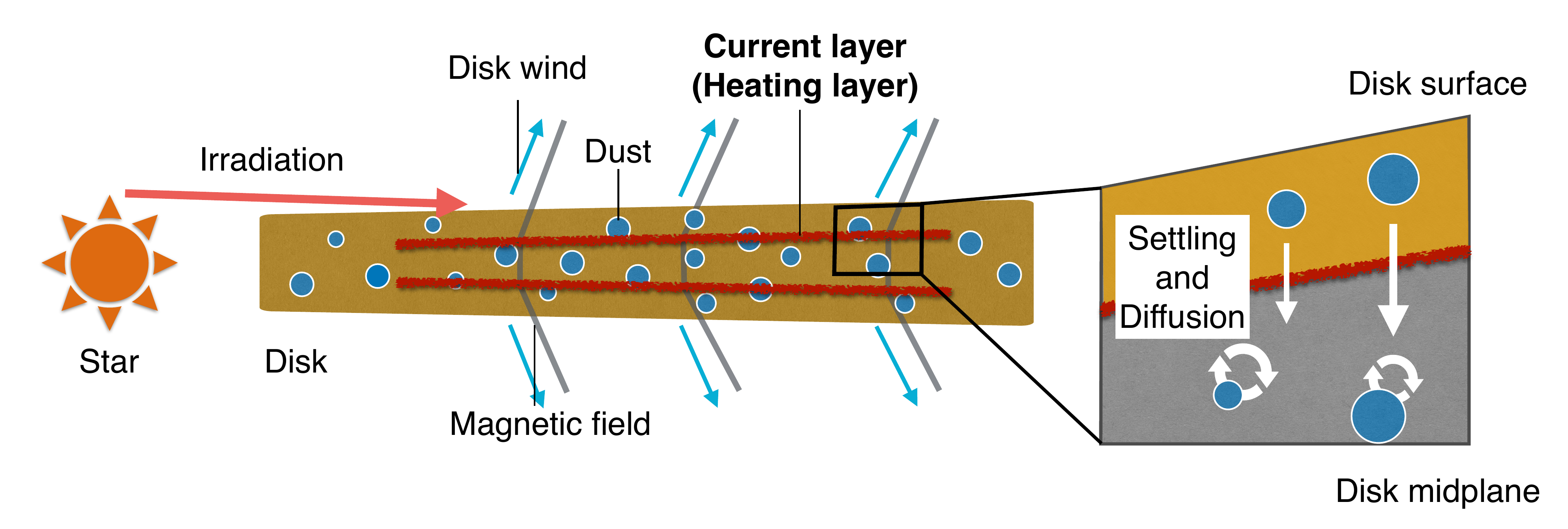}
    \includegraphics[width = 0.6\hsize,bb=0 0 1592 849]{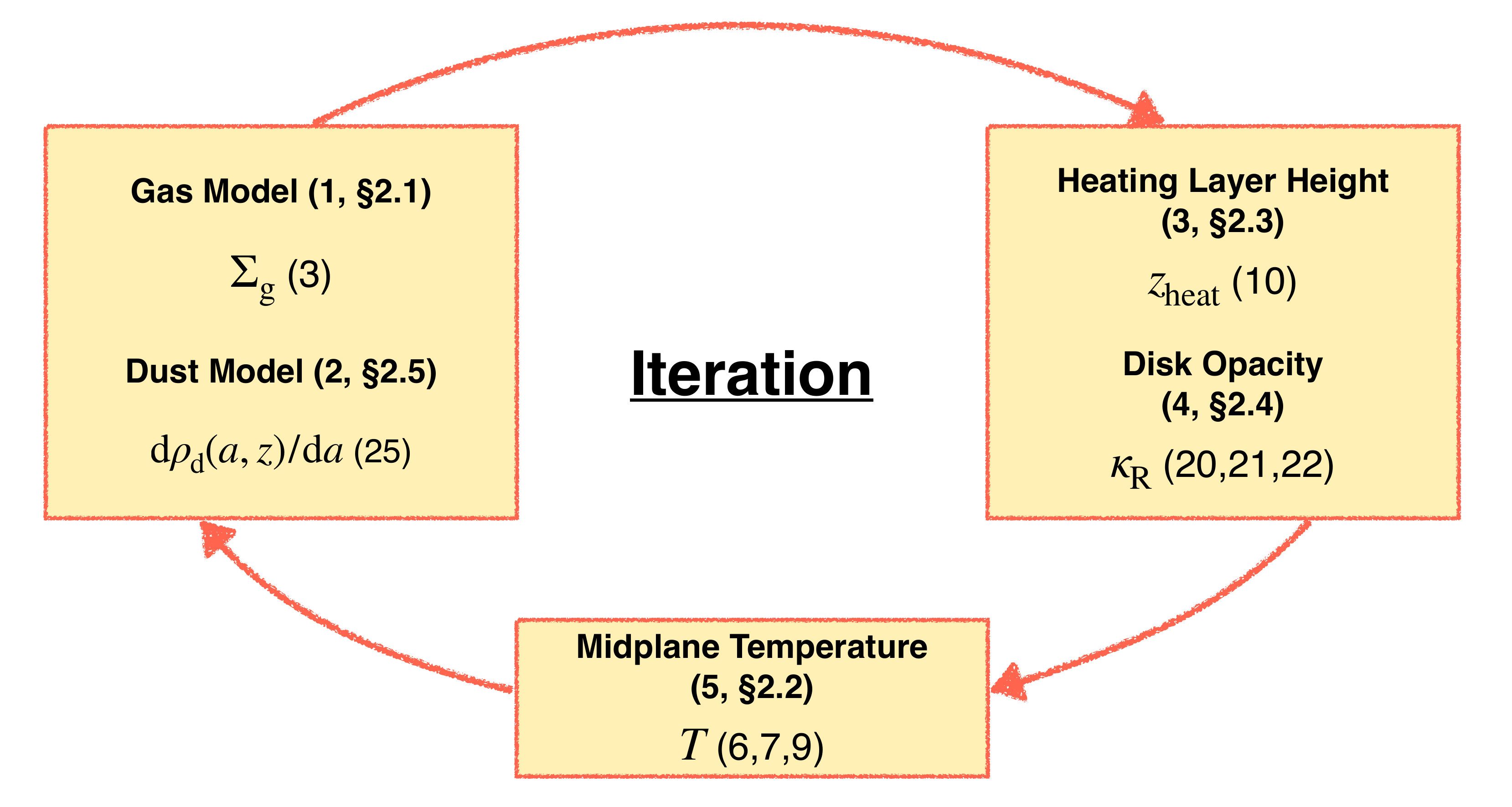}
    \caption{Schematic illustration showing the temperature model for magnetically accreting disks used in this study. The model considers a weakly ionized protoplanetary disk whose accretion is driven by magnetic winds. In addition to stellar irradiation heating, the model assumes internal Joule heating by a thin current layer lying at height $z=z_{\rm heat}$ where the ambipolar Elsasser number Am exceeds 0.3 \citep{Mori2019,Mori2021}. The model also considers weak turbulence that drives vertical dust stirring but has little effect on disk heating or accretion. Dust grains determine the radial and vertical profiles of the ionization fraction (and hence Am) and opacity. The grain vertical distribution is determined by the balance between vertical settling and diffusion. 
    The color scheme of the schematic illustration at the upper right panel indicates that the opacity above the heating layer is important in this study.
    The bottom panels show the iterative procedure that determines the midplane temperature consistently with other disk quantities (see Section~\ref{subsec:procedure}).}
    \label{fig:model}
\end{figure*}
In this section, we describe how we calculate the midplane temperature of MHD accretion disks with a distribution of dust grains (see Figure \ref{fig:model} for a schematic illustration).  
Following \citet{Mori2021}, we consider a protoplanetary disk whose accretion is driven by magnetic disk winds (Section \ref{subsec:disk}). The disk is heated by stellar irradiation and internal Joule dissipation (Section \ref{subsec:temperature}). The Joule heating occurs on a thin current layer whose height depends on the disk's ionization structure (Section \ref{sssec:current}). 

Dust grains control the disk ionization state and opacity (Sections~\ref{sssec:current} and \ref{sssec:opacity}). In this study, we consider a simple power-law grain size distribution and determine the grains' vertical distribution by assuming the presence of weak disk turbulence (Section~\ref{subsec:dust}). By weak, we mean that the turbulence causes vertical dust diffusion but gives a minor contribution to disk accretion and heating, which are dominated by laminar magnetic fields. 

Because the disk's density, ionization, opacity, and temperature structures are mutually dependent, we determine these quantities from an iterative calculation (Section~\ref{subsec:procedure}). Our model involves two important free parameters: the maximum grain size and vertical diffusion coefficient (section~\ref{subsec:parameter}). While the former determines the grains' size distribution, the latter controls their vertical distribution.

In the following, we describe the assumptions of our model in more detail.

\subsection{Gas Density Structure}\label{subsec:disk}
{We consider a disk around a solar-mass star. We assume that the disk is magnetically accreting, namely, its accretion is driven by magnetic winds. We assume that the disk is vertically nearly isothermal and in vertical hydrostatic equilibrium. Under the footpoint of disk winds, gas pressure generally dominates over magnetic pressure \citep[e.g.,][]{B&S2013b}, and therefore we neglect magnetic pressure in the vertical hydrostatic balance.}
The vertical gas density profile is then given by
\begin{equation}
    \rho_{\rm g} (z)=\frac{\Sigma_{\rm g}}{\sqrt{2\pi}H_{\rm g}}\exp \pr{-\frac{z^2}{2H_{\rm g}^2}},\label{eq:rho_g}
\end{equation}
where $z$ is the height from the disk midplane, $\Sigma_{\rm g}$ is the gas surface density, 
and $H_{\rm g}$ is the gas scale height. 
The gas scale height is 
\begin{equation}
    H_{\rm g}=\frac{\cs}{\Omega},\label{eq:h_g}
\end{equation}
where $\cs=\sqrt{k_{\rm B}T/\mu m_{\rm p}}$ is the isothermal sound speed and $\Omega=\sqrt{ G {M_\odot}/r^3}$ is the Keplerian angular velocity, 
with $k_{\rm B},\,T,\,G,$ and $r$ being the Boltzmann constant, disk temperature, gravitational constant, and radial distance from the central star, respectively. The mean molecular mass $\mu m_{\rm p}$ is the product of the mean molecular weight $\mu$ (assumed to be 2.34) and proton mass $m_{\rm p}$.
At the low altitude below the heating layer, the gas can be regarded as vertically isothermal \citep[see][]{Mori2019}. Gas above the heating layer may not be isothermal, but it would not affect our assumption. 
The optical depth of the heating layer, which determines the temperature structure, is determined by the column surface density above the heating layer. If $\rho_{\rm g}$ decreases steeply as a function of $z$, the column surface density above the heating layer is approximately determined by the amount of gas near the heating layer. Therefore, since the disk structure is not expected to change significantly even if the upper layer is non-isothermal, the assumption that the disk is vertically isothermal would be valid. 

{The gas surface density is determined by the rate of angular momentum removal by magnetic winds. For simplicity, we neglect mass removal by magnetic winds, which is equivalent to assuming that} the gas surface density $\Sigma_{\rm g}$ can then be written as \citep{Suzuki2016,Mori2021}
\begin{equation}
    \Sigma_{\rm g} =\frac{\dot{M}}{2\sqrt{2\pi}\alppz \cs r}, \label{eq:Sigma_g}
\end{equation}
where $\dot{M}$ is the disk mass accretion rate, $\alppz$ is the wind stress normalized by the midplane gas pressure (see Equation (3) of \citealt{Mori2021}).
{For simplicity, we assume that $\alppz$ is constant in space and time.}
According to MHD simulations \citep[e.g.,][]{B&S2013b,Simon13,Bai2017,Mori2019}, the value of $\alppz$ ranges between $10^{-4}$--$10^{-2}$ and depends on the net flux of the vertical magnetic field.
In this study, we take $\alppz = 
10^{-3}$ as the default value. As we show in Section~\ref{sssec:wind}, this parameter has little effect on the resulting temperature profiles.
{The $\alppz$ parameter in our model should not be confused with the $\alpha$ parameter in the standard viscous accretion model \citep{ShakuraSunyaev73}; the former represents angular momentum removal by disk winds, whereas the latter is related to radial angular momentum transfer within the disk. In terms of the efficiency of angular momentum transport, $\alpha$ is equivalent to $(r/H_{\rm g})\alppz$ \citep[see][]{Mori2021}. Therefore, an MHD disk model with $\alppz = 10^{-3}$ has a surface density comparable to that of a viscous disk with $\alpha \sim (r/H_{\rm g})\alppz \sim 0.03$, where we have used that $r/H_{\rm g} \sim  30$ at $r \sim 1~\rm au$.}

{We assume that the accretion is quasi-steady because the timescale on which the global disk accretion rate decreases (which should be comparable to {the lifetime of} $\sim \rm Myr$ of typical disks) is much longer than the timescale of local gas accretion around the snow line ( $\sim \Sigma_{\rm g}r^2/\dot{M} = r/(2\sqrt{2\pi}\alppz H_{\rm g} \Omega_{\rm K}) \sim 10^3~\rm yr$ at $r \sim 1 ~\rm au$). The accretion rate $\dot{M}$ can then be approximated as radially constant and equal to the stellar accretion rate}.
Following \cite{Mori2021}, {we model the stellar accretion rate as \citep{Hartmann2016},}
\begin{equation}
    \dot{M}= \dot{M}_{\rm 1~Myr}\pr{\frac{t}{1\,{\rm Myr}}}^{-1.07},
    \label{eq:Mdot}
\end{equation}
where $\dot{M}_{\rm 1~Myr} = 4\times 10^{-8\pm 0.5}{M_\odot\,{\rm yr^{-1}}}$ and $t$ is the stellar age defined as the time after star formation is completed, {i.e., after the end of the protostellar accretion phase and the arrival at the stellar birthline.} The variation of $\dot{M}_{\rm 1~Myr}$ reflects the scatter of the observed mass accretion rates across different sources \citep{Hartmann2016}. We use $\dot{M}_{\rm 1~Myr} = 4\times 10^{-8}{M_\odot\,{\rm yr^{-1}}}$ unless otherwise noted, but also study the effect of the variation of $ \dot{M}_{\rm 1~Myr}$ on the snow line migration in Section~\ref{sssec:amax}. Combining Equations~(\ref{eq:Sigma_g}) and (\ref{eq:Mdot}) gives
\begin{equation}
    \Sigma_{\rm g} \approx 400 \pr{\frac{t}{1\,{\rm Myr}}}^{-1.07}\pr{\frac{T}{200~\rm K}}^{-1/2} \pr{\frac{r}{1\au}}^{-1}~\rm g~cm^{-2}
    \label{eq:Sigma_g_2}
\end{equation}
for the default values $\dot{M}_{\rm 1~Myr} = 4\times 10^{-8}{M_\odot\,{\rm yr^{-1}}}$ and $\alppz = 10^{-3}$.

\subsection{Disk Midplane Temperature}\label{subsec:temperature}
Considering both stellar irradiation and accretion heating, we give the disk midplane temperature $T$ as
\begin{equation}
    T=(T_{\rm irr}^4+T_{\rm acc}^4)^{1/4},\label{eq:T}
\end{equation}
where $T_{\rm irr}$ and $T_{\rm acc}$ are the temperatures in the limits where disk heating is dominated by irradiation and accretion, respectively.

\begin{figure}
    \centering
    \includegraphics[width=\hsize, bb=0 0 282 210]{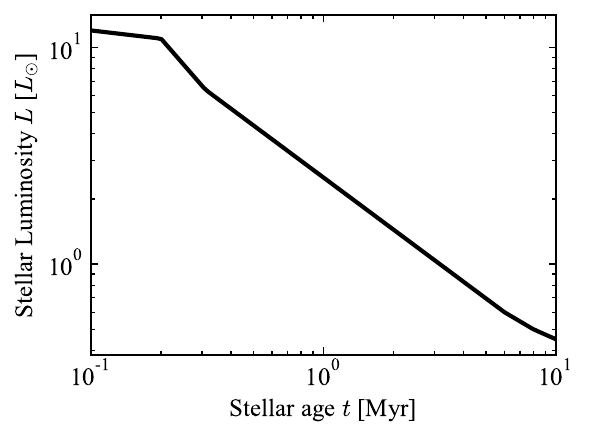}
    \caption{Stellar luminosity $L$ as a function of the stellar age $t$ from \eqref{eq:L}, which is an analytic fit to the luminosity evolution curve provided in Figure 3 of \cite{Mori2021}.
    }
    \label{fig:luminosity}
\end{figure}
The temperature in the irradiation-dominated limit is approximately given by \citep{Kusaka1970,C&G1997}
\begin{equation}
    T_{\rm irr}=110 \pr{\frac{r}{1\au}}^{-3/7}\pr{\frac{L}{{L_\odot}}}^{2/7}\K ,\label{eq:irr}
\end{equation}
where $L$ is the stellar luminosity (for details, see \citealt{Mori2021}).
\eqref{eq:irr} already uses that the central star is solar-mass.
Following \cite{Mori2021}, we consider stellar evolution and give $L$ as a function of the stellar age $t$. 
Figure 3 of \cite{Mori2021} provides $L$ versus $t$ based on the stellar evolution model of \citet{Feiden16}.  
In this study, we use {an empirical fit for} the luminosity evolution curve of \cite{Mori2021}, 
\begin{eqnarray}
    L=\left\{
    \begin{array}{ll}
        12\pr{\frac{t}{0.1\Myr}}^{{-0.13}} {L_{\odot}}, & 0.1{\Myr}\leqslant t < 0.2\Myr, \\[3mm]
        11\pr{\frac{t}{0.2\Myr}}^{{-1.2}} {L_{\odot}}, & 0.2{\Myr}\leqslant t < 0.31\Myr, \\[3mm]
        6.4\pr{\frac{t}{0.31\Myr}}^{{-0.80}} {L_{\odot}}, & 0.31{\Myr}\leqslant t < 6.0\Myr, \\[3mm]
        0.6\pr{\frac{t}{6.0\Myr}}^{{-0.63}} {L_{\odot}}, & 6.0{\Myr}\leqslant t < 8.0\Myr, \\[3mm]
        0.5\pr{\frac{t}{8.0\Myr}}^{{-0.47}} {L_{\odot}}, & 8.0{\Myr}\leqslant t \leqslant 10\Myr.
    \end{array}
    \right. \label{eq:L}
\end{eqnarray}

The midplane temperature in the accretion-dominated limit is given by \citep{Mori2021}
\begin{equation}
    T_{\rm acc}=\left[\pr{\frac{3\dot{M}\Omega^2f_{\heat}}{32\pi\sigma}}\pr{\tau_{\heat}+\frac{1}{\sqrt{3}}}\right]^{1/4},\label{eq:acc}
\end{equation}
where $\sigma$ is the Stefan--Boltzmann constant and $\tau_{\rm heat}$ is the optical depth from infinity to the heating layer.
The dimensionless parameter $f_{\rm heat}$ is the fraction of the energy converted into Joule heating in the total energy liberated by accretion. As described in Appendix B of {\cite{Mori2021}}, this fraction is the order of $10^{-3}$--1 and depends on the strength of the toroidal magnetic field within the disk. 
By default, we take $f_{\rm heat}=1$ to explore the maximum possible efficiency of MHD disk heating. 
We explain how lower $f_{\rm heat}$ affects our results in Section \ref{sssec:fdepth}.
{It may be inconsistent to consider the case $f_{\rm heat}<1$ under the assumption that mass loss by disk winds is neglected. However, because the total gas surface density $\Sigma_{\rm g}$ little affects the temperature of our disk model (see Section~\ref{sssec:alpha}), we expect that this assumption does not affect our results and conclusion.}

The optical depth to the heating layer depends on the height of the heating layer and the opacity above the {heating} layer. The former is determined by the disk's vertical ionization structure.  In the following sections, we explain how we related the disk ionization structure and opacity to the size and vertical distribution of dust grains.

\subsection{Heating Layer Height}\label{sssec:current}
We determine the heating layer height from the disk's vertical resistivity profile. 
Near the disk surface, ambipolar diffusion is the dominant nonideal MHD effect.  
The strength of ambipolar diffusion is given by the ambipolar diffusivity $\eta_{\rm A}$, which is a function of the ion and electron number densities, local gas density, and magnetic field strength. We calculate $\eta_{\rm A}$ using the generalized Ohm's law for electrons and ions \citep[e.g., Equations (25)--(31) of][]{Wardle2007}. 
{A detailed expression for $\eta_{\rm A}$ is presented in Appendix \ref{asec:eta}.}
A useful quantity to predict how strongly ambipolar diffusion affects magnetic field generation is the ambipolar Elsasser number $\Am$ defined by
\begin{equation}
    \Am =\frac{v_{\rm A}^2}{\eta_{\rm A}\Omega}, \label{eq:Am}
\end{equation}
where $v_{\rm A} = B/\sqrt{4\pi \rho_{\rm g}}$ is the \Alfven speed and $B$ is the magnetic field strength. 
The ambipolar Elsasser number is independent of $B$ when the ion and electron number densities are equal \citep[e.g.,][]{Bai2014}. 
At $r \sim 1$--$10~\rm au$, Am is mostly less than unity near the midplane and exceeds unity toward the disk surface \citep{Wardle2007,S&W2008,B&S2013a}.
MHD simulations by \citet{Mori2019} show that Joule heating mainly occurs at a layer where $\Am \sim 1$,
because no strong electric current can develop for $\Am \ll 1$   \citep[see also][]{B&S2013a,Gressel2015}
and the Joule heating associated with ambipolar diffusion is inefficient for $\Am\gg 1$.
{To take the maximum estimate of the midplane temperature, \cite{Mori2021} have chosen the critical value to be less than unity but around unity, i.e., -0.5 dex $\sim 0.3$.}
Following \cite{Mori2021}, we approximate the heating layer to be infinitesimally thin and assume that the layer lies at height $z_{\rm heat}$ where $\Am = 0.3$.
{We should note that there is some ambiguity in the choice of the critical value, but choosing the critical value of 0.3 provides a good correlation between the altitude of critical Am and the heating layer \citep[shown in Figure 12 of][]{Mori2021}.}
If $\Am > 0.3$ at all altitude, we set $z_{\rm heat} = 0$.

As shown by \cite{Mori2021}, the actual location of the heating layer also depends on the Hall effect. 
The above choice for $z_{\rm heat}$ best approximates the heating layer height when the vertical magnetic field threading the disk is in the same direction as the disk's rotation axis, for which the Hall effect amplifies toroidal magnetic fields. 
In the opposite case where the vertical magnetic field is anti-aligned with the disk rotation axis, the Hall effect acts to damp toroidal fields, pushing the current layer to a higher altitude with a lower optical depth. 
To account for this uncertainty, we write the heating layer's optical depth $\tau_{\rm heat}$ as 
\begin{equation}
    \tau_{\heat}=f_{\rm depth}\int_{z_{\heat}}^{\infty}\kappa_{\rm R}(z)\rho_{\rm g}(z)\,\d z,\label{eq:tau}
\end{equation}
where $\kappa_{\rm R}$ is the Rosseland mean opacity (per unit gas mass) for the disk's own thermal emission and $f_{\rm depth}$ is a dimensionless parameter that corrects for the actual heating layer depth.
According to Appendix C of \citet{Mori2021}, $f_{\rm depth} \sim 0.3$--1 and 0.01--0.1 for the aligned and anti-aligned cases, respectively.
In this study, we mainly consider the aligned case and take $f_{\rm depth}=1$ to study the maximum possible efficiency of MHD accretion heating. We also study the impact of lower $f_{\rm depth}$ in Section \ref{sssec:fdepth}.
The disk opacity is modeled in Section~\ref{sssec:opacity}.

To determine the altitude of $\Am=0.3$, We calculate the vertical profile of $\eta_{\rm A}$ using the disk ionization equilibrium model with charged dust \citep{Okuzumi2009}. 
In this model, the number density of charged particles is determined by the equilibrium between the ionization of neutral gas and the recombination of the charged particles in the gas phase or on the dust grains.
We represent all ion species with a single species, $\rm{HCO}^{+}$.
{The number densities of ions and electrons, $n_{\rm i}$ and $n_{\rm e}$, are given by \citep{Okuzumi2009,Okuzumi2011a}}
\begin{equation}
    n_{\rm i} = \frac{2\zeta n_{\rm g}}{ u_{\rm i} {\cal A}_{\rm tot}(1+\Psi)}\pr{1+\sqrt{1+2g(\Psi)}}^{-1},\label{eq:ion_g}
\end{equation}
\begin{equation}    
    n_{\rm e} = \frac{2\zeta n_{\rm g}}{ u_{\rm e} {\cal A}_{\rm tot}\exp(-\Psi)}\pr{1+\sqrt{1+2g(\Psi)}}^{-1},\label{eq:electron_g}
\end{equation}
where $\zeta$ and $n_{\rm g}$ are the ionization rate and the number density of the neutral gas, $u_{\rm i}$ and $u_{\rm e}$ are the mean ion and electron thermal speeds, ${\cal A}_{\rm tot}$ is the total cross-sectional area of dust per unit volume, and $\Psi$ and $g(\Psi)$ are dimensionless numbers representing the effects of grain charging and gas-phase recombination, respectively.
We assume that all charged particles which collide with dust grains are adsorbed.
The mean thermal speeds are given by $u_{\rm i(e)}=\sqrt{8k_{\rm B}T/\pi m_{\rm i(e)}}$ where $m_{\rm i(e)}$ is the ion (electron) mass.
The dimensionless number $\Psi$ is 
defined by
\begin{equation}
    \Psi\equiv -\frac{\overline{Z}e^2}{ak_{\rm B}T}, \label{eq:psi}
\end{equation}
where $a$ is the dust grain size, $e$ is the elementary charge, and $\overline{Z}e$ is the grain's mean charge.
The total dust cross section depends on the grain size distribution as 
\begin{equation}
    {\cal A}_{\rm tot}=\int_{\amin}^{\amax}\frac{\d n_{\rm d}}{\d a}\pi a^2 \d a, \label{eq:Atot}
\end{equation}
where $\d n_{\rm d}/\d a$ is the grain number density per unit grain size, and $\amin$ and $\amax$ are the minimum and maximum grain sizes, respectively.
The dimensionless quantity $g(\Psi)$ has the expression
\begin{equation}
    g(\Psi)=\frac{2K_{\rm rec}\zeta n_{\rm g}}{u_{\rm i} u_{\rm e} {\cal A}_{\rm tot}^2}\frac{\exp\Psi}{1+\Psi}, \label{eq:g_psi}
\end{equation}
where $K_{\rm rec}$ is the rate coefficient for ion--electron gas-phase recombination. For $\rm{HCO}^{+}$, one has \citep{Ganguli1988}
\begin{equation}
    K_{\rm rec}=2.4\times10^{-7}\pr{\frac{T}{300\K}}^{-0.69}\,{\rm cm}^3\,{\rm s}^{-1}. \label{eq:rec}
\end{equation}
Note that $g(\Psi)$ represents the ratio of gas-phase recombination to dust adsorption of charged particles.

From the assumption that the disk is globally neutral, we derive the equation for $\Psi$ following as \citep{Okuzumi2009,Okuzumi2011a}
\begin{equation}
    \frac{1}{1+\Psi}-\sqrt{\frac{m_{\rm e}}{m_{\rm i}}}\exp\Psi -\frac{\Psi}{\Theta} \frac{\sqrt{1+2g(\Psi)}+1}{2}=0\label{eq:eqpsi},
\end{equation}
where 
\begin{equation}
    \Theta=\frac{\zeta n_{\rm g}e^2}{{\cal A}_{\rm tot}{\cal C}_{\rm tot}k_{\rm B}T}\sqrt{\frac{\pi m_{\rm i}}{8k_{\rm B}T}},
\end{equation}
is a dimensionless quantity with 
\begin{equation}
  {{\cal C}_{\rm tot}=\int_{a_{\rm min}}^{a_{\rm max}} \dfrac{{\rm d}n_{\rm d}}{{\rm d}a}a{\rm d}a}
\end{equation}
being the total radius of dust per unit volume.
We solve \eqref{eq:eqpsi} numerically, and thus obtain $n_{\rm i}$ and $n_{\rm e}$.

The ionization rate $\zeta$ {(see Appendix \ref{asec:zeta} for detail)} includes the contributions from galactic cosmic rays \citep{U&N2009}, stellar X-rays \citep{I&G1999,B&G2009},
and radionuclides \citep{U&N2009}.

\subsection{Disk Opacity}\label{sssec:opacity}
Dust is the dominant source of disk opacity. The Rosseland mean opacity $\kappa_{\rm R}$ per gas mass is given by  
\begin{equation}
    \dfrac{1}{\kappa_{\rm R}(z)}=\dfrac{\dint_0^{\infty} \dfrac{1}{\kappa_{{\rm g}\,,\lambda}(z)}\pd{B_\lambda(T)}{T}\d\lambda}{\dint_0^{\infty}\pd{B_\lambda(T)}{T}\d\lambda},
\end{equation}
where $B_\lambda (T)$ is the Plank function and $\kappa_{{\rm g}\,,\lambda}$ is the wavelength-dependent opacity per gas mass. 
The latter is related to the dust size distribution as
\begin{equation}
    \kappa_{{\rm g}\,,\lambda}(z)=\frac{1}{\rho_{\rm g}(z)}\int_{\amin}^{\amax}\kappa_{{\rm d}\,,\lambda}(a)\frac{\d \rho_{\rm d}(a,z)}{\da}\da,\label{eq:kappa}
\end{equation}
where $\d \rho_{\rm d}/\d a = m_{\rm d} \d n_{\rm d}/\d a$ is the dust mass density per unit grain radius, $m_{\rm d}$ is the grain mass, and $\kappa_{{\rm d}\,,\lambda}$ is the opacity per grain mass. 
Our disk opacity $\kappa_{\rm R}$ depends on $z$ because of dust settling.

\begin{figure}
    \centering
    \includegraphics[width=\hsize,bb=0 0 300 213]{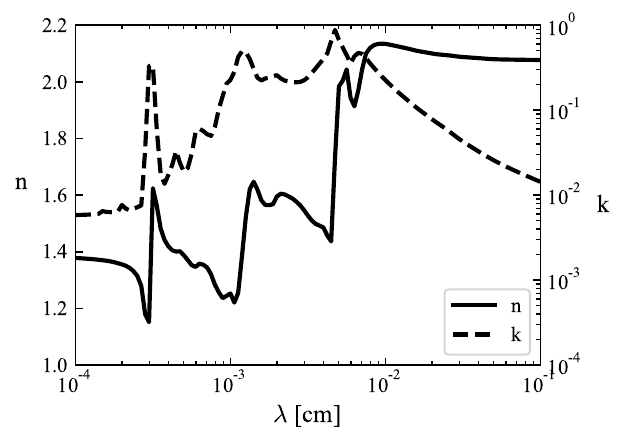}
    \caption{Real and imaginary parts of the complex refractive index, $n$ and $k$, as a function of wavelength $\lambda$ \citep{Draine2003b,W&B2008}.}
    \label{fig:nk}
\end{figure}

To compute $\kappa_{{\rm d}\,,\lambda}$, we approximate each dust grain as a uniform sphere and use the piecewise analytic formula \citep{Kataoka2014}
\begin{eqnarray}
    \kappa_{{\rm d}\,,\lambda}=\dfrac{\pi a^2}{m_{\rm d}}\times\left\{
    \begin{array}{ll}
        \dfrac{24nkx}{(n^2+2)^2}, & x \leqslant 1, \\[3mm]
        {\rm min}\pr{\dfrac{8kx}{3n}\pr{n^3-(n^2-1)^{3/2}},0.9}, & x>1,
    \end{array}
    \right. \label{eq:lam}
\end{eqnarray}
where $x=2\pi a/\lambda$ is the size parameter and
$n$ and $k$ are the real and imaginary parts of the complex refractive index, respectively.
\eqref{eq:lam} smoothly connects the expressions in the Rayleigh limit ($x \ll 1$) and geometric optics limit ($x \gg 1$). Because $\pi a^2/m_{\rm d} \propto a^{-1}$, the grain opacity is independent of $a$ and scales inversely with $a$ in the limits of $x\ll 1$ and $x \gg 1$, respectively. 
We compute $n$ and $k$ at each wavelength using the Bruggeman mixing rule, assuming that the dust around the snow line is a mixture of silicates and water ice with a mass mixing ratio of $1:1$ with no porosity.
The values of $n$ and $k$ for silicates and ice are taken from \cite{Draine2003b} and \cite{W&B2008}, respectively.
The resulting $n$ and $k$ as a function of $\lambda$ are shown in \figref{fig:nk}.
The internal density of the dust grains is $\rho_{\rm int}=1.46\gcmcmcm$.
As discussed in Section~\ref{subsec:porosity}, the results presented in Section~\ref{sec:result} would be applicable to porous grains if we replace $\amax$ with the product of $\amax$ and the grains' filling factor.
To check how strongly our results depend on the assumed dust composition, we have also conducted calculations assuming pure silicate grains with $\rho_{\rm int} = 3.5 \gcmcmcm$. The results show that assuming pure silicates instead of silicate--ice mixture only decreases the snow line's arrival time at 1 au by $\sim 0.4$ Myr at most. This impact is small compared to that of the uncertainty in the accretion rate (see Figure~\ref{fig:time_snow} in Section~\ref{sssec:amax}).  

\subsection{Dust Size and Vertical Distribution}\label{subsec:dust}
The model formulation presented above does not rely on any particular form of grain size and vertical distribution. To explore the impacts of dust growth on the snow line evolution quantitatively, we adopt a simple power-law size distribution 
\begin{equation}
    \frac{\d\Sigma_{\rm d}(a)}{\da}=\frac{\Sigma_{\rm d}}{2(\sqrt{\amax}-\sqrt{\amin})}a^{-1/2},\label{eq:size}
    \label{eq:dSigmad_da}
\end{equation}
where $\d\Sigma_{\rm d}(a)/\da$ is the dust surface density per unit grain radius and
\begin{equation}
    \Sigma_{\rm d}=\int_{\amin}^{\amax} \frac{\d \Sigma_{\rm d}(a)}{\da}\da \label{eq:Sigma_d}
\end{equation}
is the total dust surface density.
The assumed power-law slope $-1/2$ of the size distribution follows the size distribution of interstellar dust \citep{Mathis1977}, for which the grain number density per unit grain size scales as $a^
{-7/2}$.
We take $\amax$ as a free parameter and investigate the effects of the grain size distribution on MHD disk heating.
We set $\amin =0.1\um$ because smaller particles grow quickly through Brownian motion \citep{Birnstiel2011}.

We assume that $\Sigma_{\rm d}$ scales with $\Sigma_{\rm g}$ and take the default value of the dust-to-gas surface density ratio $f_{\rm dg} = \Sigma_{\rm d}/\Sigma_{\rm g}$ to be the interstellar value $f_{\rm dg} = 0.01$ {\citep{Bohlin+1978}}.
In reality, dust growth to $\amax \gg 1~\micron$ would affect the value of $f_{\rm dg}$ because large grains tend to drift toward the central star rapidly \citep{Whipple72,Adachi1976,Weidenschilling77a}. The dust-to-gas ratio in the inner disk region can either increase or decrease depending on whether the loss of the dust in the inner region is slower than the replenishment of dust drifting in from the outer region \citep[e.g.,][]{Birnstiel12}.  
In Section \ref{sssec:fdg}, we vary $f_{\rm dg}$ to 0.1 and $10^{-3}$ and study how the increase and decrease of $f_{\rm dg}$ affect our results. The scenario $f_{\rm dg} = 0.1$ is less likely because such a high dust-to-gas surface density ratio can trigger the streaming instability, which would convert the dust into planetesimals, in particular when $a_{\rm max}$ is large and {vertical diffusion is weak \citep[e.g.,][]{Johansen09,Yang17}.} 

To determine the vertical variation of $\d\rho_{\rm d}/\da$, we assume the balance between vertical dust settling and diffusion.
This balance gives \citep{T&L2002}
\begin{equation}
    \frac{\d\rho_{\rm d}}{\da}=\frac{\d\Sigma_{\rm d}}{\da}\cdot C_{\rm d}\exp \left[-\frac{z^2}{2H_{\rm g}^2}-\frac{\rm St_{\rm mid}}{\alpha_{\rm D}}\pr{\exp \pr{\frac{z^2}{2H_{\rm g}^2}}-1}\right], \label{eq:rho_d}
\end{equation}
where $\alpha_{\rm D}$ is a dimensionless parameter that characterizes the level of the vertical diffusion, 
${\rm St_{\rm mid}}$ is the Stokes number at the midplane for grains with size $a$, and 
$C_{\rm d}$ is the normalization constant determined by the relation ${\d \Sigma_{\rm d}}/{\d a} = \int ({\d \rho_{\rm d}}/{\d a})\dz$.
{Since most dust grains lie at $z\ll H_{\rm g}$, the exponential factor in $\d\rho_{\rm d}/\da$ can be approximated by $\exp[-z^2/(2H_{\rm d}^2)]$ when computing ${\d \Sigma_{\rm d}}/{\d a}$, yielding \citep{Fukuhara2021}}
\begin{equation}
    C_{\rm d}=\frac{1}{\sqrt{2\pi}H_{\rm d}} {,}
\end{equation}
where 
\begin{equation}
    H_{\rm d}=\pr{1+\frac{\rm St_{\rm mid}}{\alpha_{\rm D}}}^{-1/2}H_{\rm g}{,} \label{eq:H_d}
\end{equation}
is the grain scale height around the midplane.
Assuming that the grains are smaller than the mean free path of the disk gas molecules \citep[$ a\la 1~\rm cm$ at $r \sim 1~\rm au$; e.g.,][]{Adachi1976}, the midplane Stokes number has the simple expression \citep{Birnstiel2010}
\begin{equation}
    {\rm St_{\rm mid}}=\frac{\pi\rhoint a}{2\Sigma_{\rm g}}. \label{eq:St}
\end{equation}

\eqref{eq:H_d} indicates that grains with ${\rm St}_{\rm max} > \alpha_{\rm D}$ experience a high degree of settling. 
Substituting Equation~(\ref{eq:Sigma_g_2}) into \eqref{eq:St} gives 
\begin{equation}
    {\rm St_{\rm mid}}\approx 6\times 10^{-6} \pr{\frac{t}{1\,{\rm Myr}}}^{1.07} \pr{\frac{T}{170\,\K}}^{1/2} \pr{\frac{r}{1\au}} \pr{\frac{a}{10\um}}{,} \label{eq:St_a}
\end{equation}
for $\dot{M}_{\rm 1~Myr} = 4\times 10^{-8}\Mpyr$, $\alppz = 10^{-3}$, and $\rho_{\rm int}$ given in Section~\ref{sssec:opacity}.
Therefore, for $t \sim 1~\rm Myr$, $T \sim 10^2~\rm K$, $\alppz \sim 10^{-3}$, and $r \sim 1~\rm au$, the condition ${\rm St}_{\rm mid} \ga \alpha_{\rm D}$   reduces to $a \ga  100(\alpha_{\rm D}/10^{-4})~\micron$.

\subsection{Iterative Procedure} \label{subsec:procedure}
To calculate the disk temperature $T$ with other disk quantities, we perform the following iterative calculation at each time $t$ and radial location $r$ (see the bottom panels of Figure~\ref{fig:model}). 
\begin{enumerate} 
    \item For a given mass accretion rate, calculate the vertical gas density distribution $\rho_{\rm g}(z)$ (Section \ref{subsec:disk}). 
    
    \item For given $\amax$ and $\alpha_{\rm D}$, calculate the dust size and spatial distribution $\d\rho_{\rm d}(a,z)/\da$ (Section~\ref{subsec:dust}).
    
    \item Using $\d\rho_{\rm d}/\da$, calculate the disk ionization structure and determine the heating layer height $z_{\rm heat}$ (Section \ref{sssec:current}).
   
    \item Also calculate the vertical distribution of the disk opacity $\kappa_{\rm R}(z)$ (Section \ref{sssec:opacity}).

    \item Using $z_{\rm heat}$ and $\kappa_{\rm R}(z)$, calculate the heating layer optical depth $\tau_{\rm heat}$ \eqref{eq:tau} and the contribution to the disk temperature from internal accretion heating, $T_{\rm acc}$ \eqref{eq:acc}. Obtain the final midplane temperature $T$ by adding the contribution from stellar irradiation heating {(Equations~(\ref{eq:T}) and (\ref{eq:irr}))}.
    
\end{enumerate}
For every set of $t$ and $r$, we repeat this procedure until $T$ converges to a precision of $10^{-3}$. Our calculations show that the final temperature is independent of the initial temperature given at the beginning of the iteration. 

\subsection{Parameter Choices}\label{subsec:parameter}

The maximum dust size $\amax$ and vertical diffusion coefficient $\alpha_{\rm D}$ are key parameters of our model.
We vary $\amax$ between  $0.11\um$--$1.0\cm$ to examine the effects of grain growth on the disk temperature.
The choice $\amax=0.11\um$ yields nearly monodisperse grain size distribution peaked at $a\approx 0.1\um$ and corresponds to the choice of  \citet{Mori2021}.

The source of vertical diffusion around the snow line is highly uncertain. Because magnetic turbulence is likely suppressed {\citep{Gammie1996,B&S2013a,Gressel2015}}, only non-MHD turbulence would produce $\alpha_{\rm D}$.
{Potential drivers of turbulence around the snow line include the vertical shear instability \citep{UrpinBrandenburg98,Nelson+13,LinYoudin15}, convective overstability \citep{K&H2014,Lyra2014}, and zombie vortex instability \citep{Marcus2015,Marcus2016,Barranco2018}, which are hydrodynamical instabilities that operate depending on the disk's thermal relaxation timescale \citep{L&U2019,Malygin2017,P&K2019}.
As discussed in Section~\ref{subsec:alpha}, the vertical shear instability and convective overstability can potentially produce turbulence with $\alpha_{\rm D} \sim 10^{-5}$--$10^{-3}$ around the snow line.
For this reason,} we vary $\alpha_{\rm D}$ between $10^{-5}$--$10^{-3}$.

\section{Results}\label{sec:result}
In this section, we use the model described in Section \ref{sec:model} to study how the dust size distribution affects the disk thermal structure and snow line evolution.  Section \ref{subsec:iteration} presents the dependence of the disk vertical structure on $\amax$, and Section \ref{subsec:snow} presents the evolution of the snow line for different values of $\amax$. 

\subsection{Disk Thermal Structure}\label{subsec:iteration}
To begin with, we select the particular case of $t=0.6{\Myr}$ and $\alpha_{\rm D}=10^{-3}$ and study how the disk's thermal structure depends on the maximum grain size $a_{\rm max}$. For this value of $\alpha_{\rm D}$, grains with $a \la 1~\rm mm$ are well mixed vertically (see Section~\ref{subsec:dust}) so that dust settling little affects the disk temperature structure.

\begin{figure}
    \centering
    \includegraphics[width=\hsize,clip,bb=0 0 282 282]{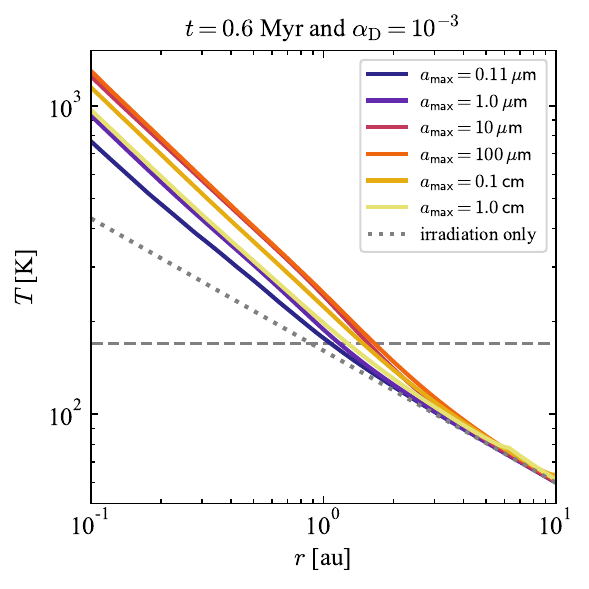}
    \caption{
    Radial profiles of the midplane temperature $T$ for $t=0.6{\Myr}$ and $\alpha_{\rm D}=10^{-3}$ with different values of $\amax$. The dotted line shows the midplane temperature for the irradiation-dominated disk. The dashed horizontal line marks the water ice sublimation temperature $T=170\K$ assumed in this study.
    }
    \label{fig:T_disk}
\end{figure}
\figref{fig:T_disk} shows the radial profiles of $T$ for different values of $\amax$. Overall, accretion heating dominates at $r\lesssim 3\au$. Importantly, we find that the profile depends on $\amax$ non-monotonically, with $\amax \sim 10$--100 $\micron$ giving the highest temperatures. In the following, we explain this behavior by looking at the disk vertical structure at $r = 1~\rm au$ in more detail.
{In Appendix \ref{asec:T_ev}, Figure~\ref{fig:T_ev} shows the radial profiles of $T$ at different times for $a_{\rm max}=0.11\um$, $a_{\rm max}=10\um$, and $a_{\rm max}=100\um$.}

\begin{figure}
    \centering
    \includegraphics[width=\hsize,bb=0 0 278 225]{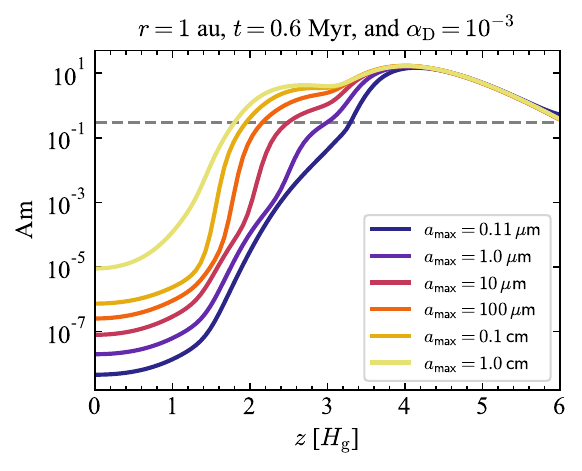}
    \caption{Vertical profiles of the ambipolar Elsasser number Am for different maximum grain sizes $a_{\rm max}$ for $r=1\au$, $t=0.6{\Myr}$, and $\alpha_{\rm D}=10^{-3}$. The gray dashed line indicates the critical Elsasser number, $\Am=0.3$, at which a strong current layer develops.
    }
    \label{fig:Am}
\end{figure}
\begin{figure}
    \centering
    \includegraphics[width=\hsize, bb=0 0 286 225]{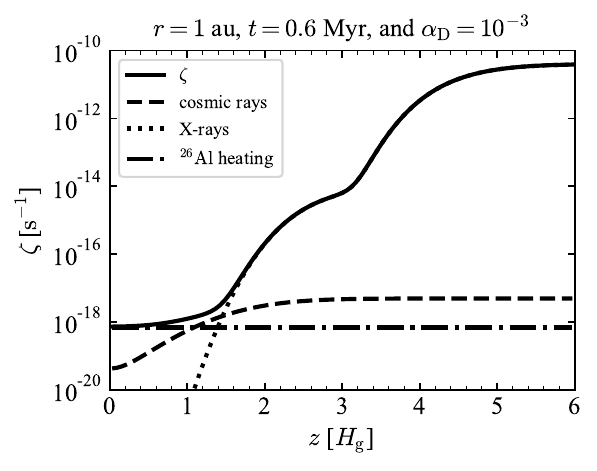}
    \caption{Vertical profile of the disk ionization rate $\zeta$ (solid line) for $r=1\au$, $t=0.6{\Myr}$, and $\alpha_{\rm D}=10^{-3}$. The dashed, dotted, and dot-dashed lines show the contributions of cosmic rays, stellar X-rays, and $^{26}{\rm Al}$ decay, respectively.}
    \label{fig:zeta}
\end{figure}
{We recall that the disk internal temperature increases as the heating layer height $z_{\rm heat}$ decreases or the opacity above the heating layer increases (Equations~(\ref{eq:acc}) and (\ref{eq:tau})).
Of the two, $z_{\rm heat}$ is determined by the vertical profile of the ambipolar Elsasser number Am} (Section \ref{sssec:current}). \figref{fig:Am} shows the vertical profiles of Am for different values of $a_{\rm max}$. 
Overall, Am is well below unity at the midplane and exceeds the critical value $\Am = 0.3$ at $z \sim 2$--$3H_{\rm g}$. Around these heights, the X-rays are the dominant ionizing sources as shown in \figref{fig:zeta}. {The steep decrease of Am at $z \la 2H_{\rm g}$ reflects the decrease of the X-ray ionization rate at the high column depths due to the attenuation of scattered X-rays \citep{I&G1999,B&G2009}.}

\begin{figure*}
    \centering
    \includegraphics[width=0.9\hsize,bb=0 0 576 425]{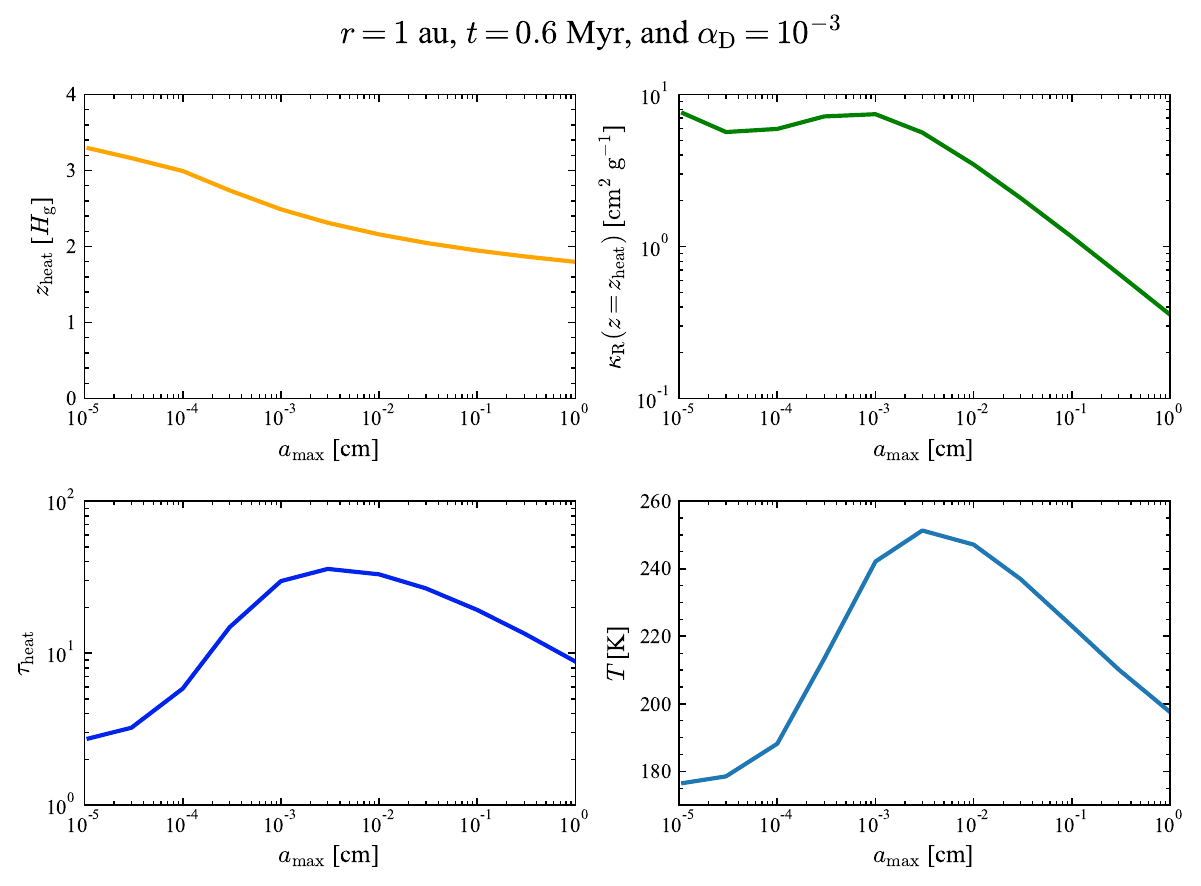}
    \caption{
    Key quantities characterizing the disk thermal structure as a function of the maximum grain size $\amax$ for $r=1\au$, $t=0.6{\Myr}$, and $\alpha_{\rm D}=10^{-3}$. The upper left, upper right, lower left, and lower right panels show the heating layer height $z_{\rm heat}$, opacity $\kappa_{\rm R}$ at $z = z_{\rm heat}$, heating layer optical depth $\tau_{\rm heat}$, and midplane temperature $T$, respectively. 
    }
    \label{fig:amax}
\end{figure*}

The upper left panel of \figref{fig:amax} shows $z_{\rm heat}$ as a function of $a_{\rm max}$. We find that $z_{\rm heat}$ decreases monotonically with $\amax$.
To understand how $z_{\rm heat}$ depends on $a_{\rm max}$, we first note that the variation of Am only comes from those of $n_i$ and $\Omega$ when $n_{\rm i} \approx n_{\rm e}$ \citep[e.g.,][]{C&M2007}. In our calculations for $f_{\rm dg} \leq 0.01$, the condition $n_i \approx n_e$ is fulfilled at $z \sim z_{\rm heat}$. 
We also note that at $z \sim z_{\rm heat}$, the adsorption of electrons and ions onto dust grains dominates over gas-phase recombination, i.e., $g(\Psi)\ll1$. In this limit, the expression for $n_{\rm i}$ can be simplified as 
\begin{equation}
    n_{\rm i} = \frac{\zeta n_{\rm g}}{u_{\rm i} {\cal A}_{\rm tot}}\frac{1}{1+\Psi},\label{eq:ion}
\end{equation}
with $\Psi$ being constant for $n_i \approx n_e$ \citep[see][]{Okuzumi2009,Okuzumi2011a}. 
Therefore, the dependence of $n_{\rm i}$ on $a_{\rm max}$ only comes from that of the grains' total geometric cross section ${\cal A}_{\rm tot}$.
When $\d\Sigma_{\rm d}/\d a \propto a^{-1/2}$, the smallest grains dominate the total cross section because $\int_{a_{\rm min}}^{a_{\rm max}} (\pi a^2 /m_{\rm d})  a^{-1/2} \d a \propto \int_{a_{\rm min}}^{a_{\rm max}} a^{-3/2} \d a \propto a_{\rm min}^{-1/2}$ for $a_{\rm min} \ll a_{\rm max}$. The size of the largest grains still matters because the fraction of the smallest grains decreases with increasing $a_{\rm max}$, the effect encapsulated in the prefactor $\Sigma_{\rm d}/(\sqrt{a_{\rm max}}- \sqrt{a_{\rm min}}) \approx \Sigma_{\rm d}/\sqrt{a_{\rm max}}$ in the definition of $\d\Sigma_{\rm d}/\d a$ (\eqref{eq:dSigmad_da}).   Thus, a larger $\amax$ leads to a higher ${\rm Am}(z)$ and hence to a lower $z_{\rm heat}$. This conclusion is little affected by dust settling because the smallest grains are well mixed vertically. 
{At large $\amax$,  $z_{\rm heat}$ stays around $2H_{\rm g}$ because $\zeta$ and Am decrease steeply below this height (note that $z_{\rm heat}$ is defined as the height where Am crosses a certain value).}

The upper right panel of \figref{fig:amax} shows how the disk opacity $\kappa_{\rm R}$ at the heating layer varies with $a_{\rm max}$. One can see that the opacity is constant for $\amax \la 10\um$ and decreases with increasing $\amax$ for $\amax \ga 10\um$. This threshold grain size corresponds to the Planck function's peak wavelength $\lambda_{\rm peak}$, which is $\sim 10~\rm \mu m$ at $T \sim 100~\rm K$. In our opacity model, the Rosseland mean opacity given for $T$ is mainly determined by the grain monochromatic opacities around $\lambda \sim \lambda_{\rm peak}$. 
For $\amax\lesssim \lambda_{\rm peak}$, most grains fall into the Rayleigh limit and provide equal $\kappa_{\rm d,\, \lambda_{\rm peak}}$. For $\amax\gtrsim \lambda_{\rm peak}$, grains with $a \gtrsim \lambda_{\rm peak}$ have lower $\kappa_{\rm d,\, \lambda_{\rm peak}}$ ($ \propto a^{-1}$), and therefore $\kappa_{\rm R}$ decreases with increasing $a_{\rm max}$. The upper right panel of \figref{fig:amax} shows that the opacity in the latter regime scales with $a_{\rm max}^{-1/2}$. This scaling can be derived analytically by using $\kappa_{\rm R} \sim \kappa_{\rm g,\, \lambda_{\rm peak}}$, neglecting dust settling ($\d\rho_{\rm d}/\d a \propto \d\Sigma_{\rm d}/\d a \propto a^{-1/2}$), and approximating $\kappa_{\rm d, \lambda_{\rm peak}} \propto a^{0}$ and $a^{-1}$ for $a < \lambda_{\rm peak}$ and $a > \lambda_{\rm peak}$, respectively.

\begin{figure*}
    \centering
    \includegraphics[width=\hsize,clip,bb=0 0 1920 580]{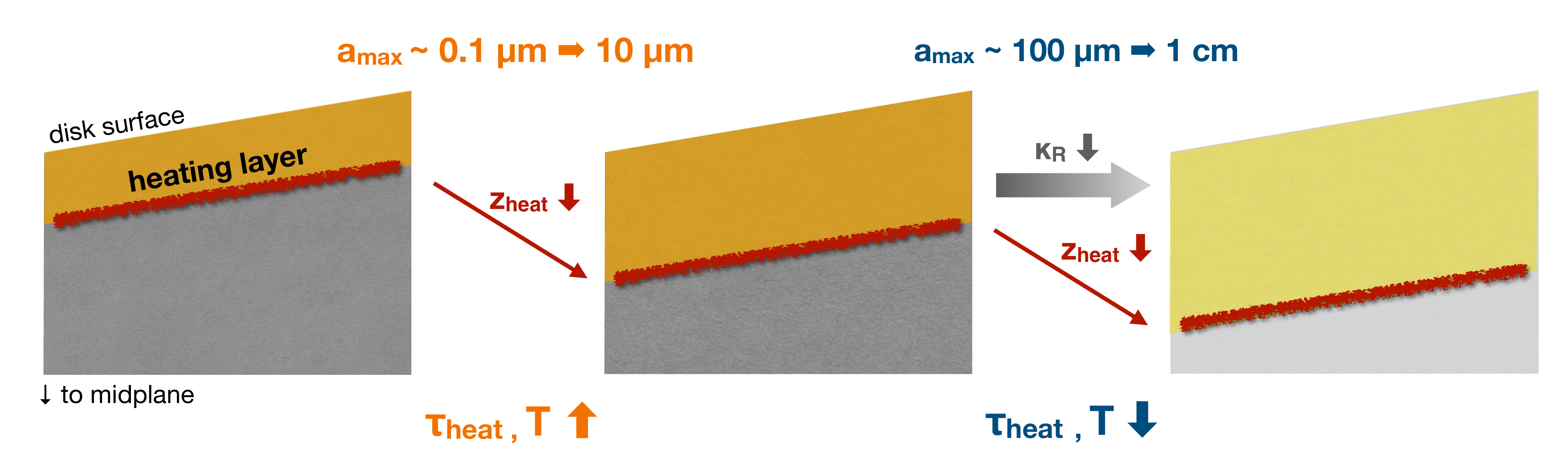}
    \caption{
    Schematic illustration showing how the variation of the heating layer height $z_{\rm heat}$ and opacity $\kappa_{\rm R}$ with the maximum grain size $\amax$ affects the disk internal temperature $T$ {in the fiducial case (Section~\ref{sssec:amax})}. As  $\amax$ increases, the ionization fraction increases, and hence  $z_{\rm heat}$ decreases (this holds unless $f_{\rm dg}$ is well above 0.01; see Section~\ref{sssec:fdg}). For $\amax \la 10\um$, $\kappa_{\rm R}$ is approximately constant, and therefore increasing $\amax$ results in increasing optical depth above the heating layer, $\tau_{\rm heat}$, and consequently in increasing $T$. 
    For $\amax \ga 100\um$, the decrease of $\kappa_{\rm R}$ causes the decrease of $\tau_{\rm heat}$, resulting in decreasing $T$ with increasing $\amax$. {This picture applies to more general cases where the decrease of $\kappa_{\rm R}$ dominates over the decrease of $z_{\rm heat}$; the case with $f_{\rm dg}= 0.1$ represents one exception (see Section~\ref{sssec:fdg}).} 
    }
    \label{fig:10-100um}
\end{figure*}

The different $\amax$ dependences of $z_{\rm heat}$ and $\kappa_{\rm R}$ are key to understanding the non-monotonic $\amax$ dependence of $T$ shown in Figure~\ref{fig:T_disk}.
The bottom panels of \figref{fig:amax} show the heating layer optical depth $\tau_{\rm heat}$ and midplane temperature $T$ as a function of $\amax$. We find that $\tau_{\heat}$ and $T$ are maximized at $\amax\sim 10$--$100\um$. 
\figref{fig:10-100um} schematically shows why this range of $\amax$ maximizes the efficiency of disk internal heating {for the fiducial case}. As shown in the upper left panel of \figref{fig:amax}, the heating layer height decreases monotonically with increasing $\amax$. Therefore, a larger $\amax$ always leads to a larger column surface density above the heating layer. At $\amax \la 10~\rm \mu m$, the opacity around the heating layer is constant (see the upper right panel of \figref{fig:amax}), and therefore a larger $\amax$ results in a larger $\tau_{\rm heat}$ and in turn a larger $T$ (see Equations~(\ref{eq:acc}) and (\ref{eq:tau})). 
In contrast, at $\amax \ga 100~\rm \mu m$, the opacity decreases with increasing $\amax$. {In this example, the decrease of $z_{\rm heat}$ at $\amax \ga 100~\rm \mu m$ is rather slow for the reason already explained. The slower increase of the column depth above $z=z_{\rm heat}$ than the decrease of $\kappa_{\rm R}$ results in the decrease of $\tau_{\rm heat}$ and $T$ beyond $\amax \sim 100~\rm \mu m$. This picture applies as long as $z_{\rm heat}$ for $\amax \ga 100~\rm \mu m$ is determined by the steep drop in $\zeta$ below $z \approx 2H_{\rm g}$. One exceptional case is presented in Section~\ref{sssec:fdg}.}

\subsection{Snow Line Evolution}\label{subsec:snow}
We now use the disk temperature profiles at different times to derive the temporal evolution of the snow line for various model parameters. For the sale of convenience, we approximate the water sublimation temperature to $170\K$. The water snow line thus corresponds to the location where  $T=170\K$. 
We begin by showing the results for $\alpha_{\rm D}=10^{-3}$ in Section \ref{sssec:amax}, and then explore the dependence of the results on our parameters. We show the dependence on $\alppz$, $\alpha_{\rm D}$, $f_{\rm dg}$, and $f_{\rm heat}$ and $f_{\rm depth}$ in Sections~\ref{sssec:wind}, \ref{sssec:alpha}, \ref{sssec:fdg}, and \ref{sssec:fdepth}, respectively.

\subsubsection{Fiducial Case}\label{sssec:amax}

\begin{figure}
    \centering
    \includegraphics[width=\hsize, bb=0 0 284 282]{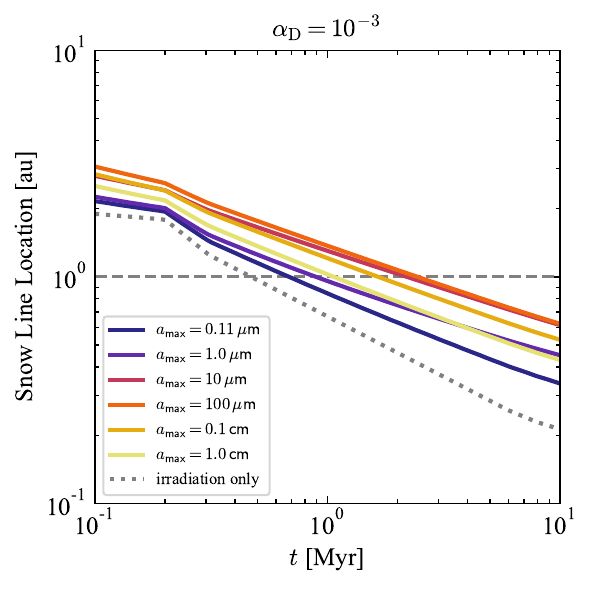}
    \caption{
    Snow line radius as a function of the stellar age $t$ for different values of the maximum grain size $\amax$ with $\alpha_{\rm D}=10^{-3}$. The dotted line is for the irradiation-dominated limit. The dashed horizontal line marks the current Earth's orbital radius, $1\au$.
    }
    \label{fig:snow}
\end{figure}
\figref{fig:snow} shows the snow line radius as a function of time $t$ for different values of $\amax$ with $\alpha_{\rm D} = 10^{-3}$. In general, the snow line migrates inward because both the disk accretion rate $\dot{M}$ and luminosity $L$ decrease with increasing $t$ (see Equations (\ref{eq:Mdot}) and (\ref{eq:L})). For fixed $t$, the snow line radius is the largest for $\amax \sim 10$--100 $\rm \mu m$ because $\amax$ in this range maximizes the efficiency of accretion heating (see Section~\ref{subsec:iteration}).

\begin{figure}
    \centering
    \includegraphics[width=\hsize,clip,bb=0 0 272 228]{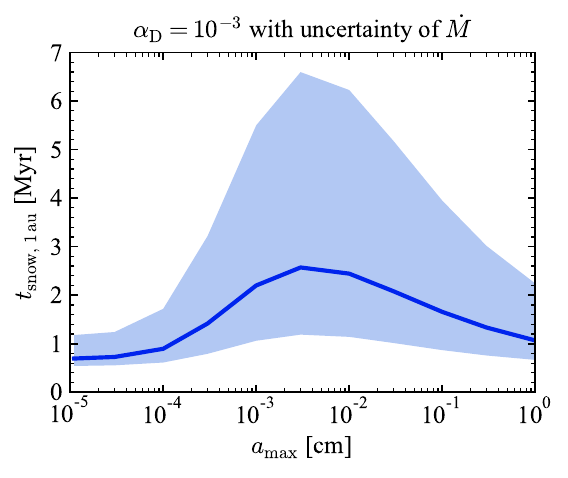}
    \caption{
    Snow line's arrival time at $1\au$ $t_{\rm snow,\,1\au}$ as a function of $\amax$ for $\alpha_{\rm D}=10^{-3}$. The shaded area represents the observational uncertainty of the mass accretion rate.
    }
    \label{fig:time_snow}
\end{figure}
We are interested in the time at which the snow line passes the current Earth's orbit of $r = 1~\rm au$. In the following, we refer to this time as $t_{\rm snow, 1\au}$. The solid line in \figref{fig:time_snow} shows $t_{\rm snow, 1\au}$ as a function of $\amax$ for the $\alpha_{\rm D}=10^{-3}$ disk model.
For $\amax = 0.11~\rm \mu m$, the snow line arrives at 1 au within $t = 1$ Myr, consistent with the finding by \cite{Mori2021}. Our new model shows that the snow line's arrival at $1\au$ can be delayed up to $t \sim$ {2--3} $\Myr$ if $\amax$ lies in the range of 10--100 $\rm \mu m$. We also emphasize that dust growth to $a_\mathrm{max} \sim 1$ mm still substantially delays the snow line migration to $\sim 2$ Myr. Therefore, dust growth is an important factor that determines the location and evolution of the snow line.
We discuss $a_{\rm max}$ determined by dust growth and planetesimal formation in protoplanetary disks in Section~\ref{subsec:planetesimal}.

So far, we have fixed the mass accretion rate at $1 M_{\odot}\,{\rm yr}^{-1}$, $\dot{M}_{\rm 1\Myr}$, to be $4 \times 10^{-8} M_{\odot}\,{\rm yr}^{-1}$. As mentioned in Section~\ref{subsec:disk}, the observationally determined accretion rates of pre-main sequence stars have a scatter of $\sim 0.5~\rm dex$. The impact of this observational scatter is also shown in Figure~\ref{fig:time_snow}, where the upper and lower boundaries of the shaded area represent $t_{\rm snow, 1\au}$ for $\dot{M}_{\rm 1\Myr} =4 \times 10^{-8+0.5}$ and $4\times 10^{-8-0.5}M_{\odot}\,{\rm yr}^{-1}$, respectively. 
The scatter of $t_{\rm snow, 1\au}$ is the largest at $\amax=10$--$100\um$, where the accretion heating is the most efficient. The figure indicates that $t_{\rm snow, 1\au}$ can be as large as $\sim 6\Myr$ if $\dot{M}_{\rm 1\Myr}$ is at the high end of the scatter.

\subsubsection{Dependence on $\alppz$}\label{sssec:wind}

\begin{figure}
    \centering
    \includegraphics[width=\hsize, bb=0 0 281 227]{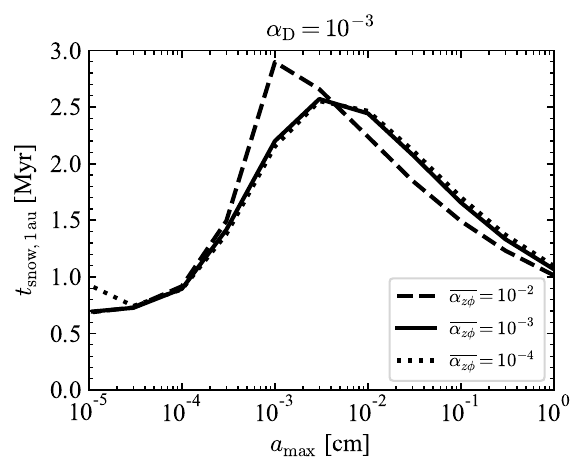}
    \caption{
    Snow line's arrival time at $1\au$ as a function of $\amax$ for different values of $\alppz$ with $\alpha_{\rm D}=10^{-3}$.
    }
    \label{fig:wind}
\end{figure}
Because we consider wind-driven accretion, the gas surface density $\Sigma_{\rm g}$ is inversely proportional to the assumed level of wind stress, which is characterized by $\alppz$ (see \eqref{eq:Sigma_g}). 
{However, this parameter little affects the temperature distribution of our disk model because it is the column density above the heating layer, not the total column density, that determines $\tau_{\rm heat}$ and $T_{\rm acc}$ (see Equations (\ref{eq:acc}) and (\ref{eq:tau})).} 
{As the disk's total surface density decreases, X-rays can penetrate more deeply into the interior and the heating layer approaches the midplane. Since this is determined by the penetration length of X-rays, the upper column density evolves while maintaining a nearly constant value. Therefore, the total surface density is not important for $\tau_{\rm heat}$, and $\alppz$ little affects $T_{\rm acc}$.}
This was already demonstrated by \citet{Mori2021} in the particular case of $\amax = 0.1~\micron$. 
Here, we show that this is also the case for larger values of $\amax$. 

Figure~\ref{fig:wind} shows $t_{\rm snow, 1\au}$ as a function of $\amax$ for different values of $\alppz$. 
One can see that $t_{\rm snow, 1\au}$ is insensitive to $\alppz$ for most cases. This is because the column density above the heating layer tends to be determined by the penetration depth of the ionizing X-rays \citep{Mori2021}. The only notable effect of varying $\alppz$ is the increase of $t_{\rm snow, 1\au}$ at $\amax \sim 10~\micron$ when we increase $\alppz$ to $10^{-2}$. In this case, the heating layer reaches the midplane and its column depth {becomes} half the total surface density before the snow line arrives at $1\au$ (Section \ref{sssec:current}).
However, this effect is minor compared with the effects of varying other parameters as we show below.

\subsubsection{Dependence on $\alpha_{\rm D}$}\label{sssec:alpha}

\begin{figure}
    \centering
    \includegraphics[width=\hsize, bb=0 0 281 225]{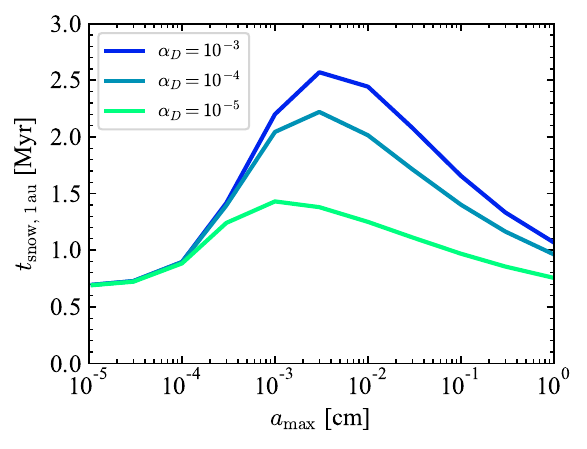}
    \includegraphics[width=\hsize, bb=0 0 287 225]{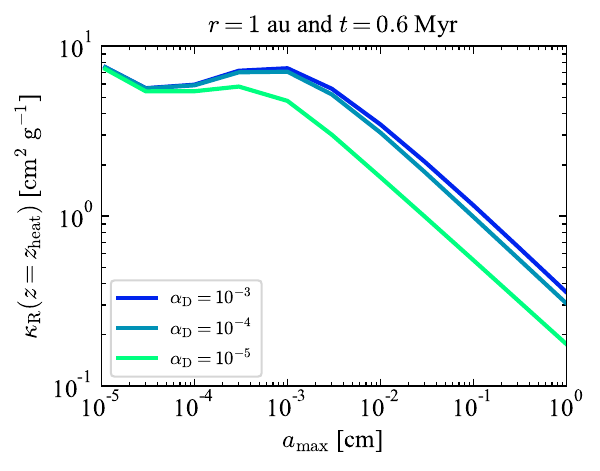}        
    \caption{
    Snow line's arrival time at $1\au$ (upper panel) and disk opacity $\kappa_{\rm R}$ at $z=z_{\rm heat}$, $r = 1~\rm au$, and $t = 0.6\Myr$  (lower panel) as a function of $\amax$ for different levels of vertical diffusion $\alpha_{\rm D}$.
    }
    \label{fig:alpha}
\end{figure}
Here, we examine how the level of vertical diffusion, $\alpha_{\rm D}$, affects the snow line migration. The upper panel of \figref{fig:alpha} shows $t_{\rm snow,\, 1\au}$ as a function of $\amax$ for different values of $\alpha_{\rm D}$. 
While $t_{\rm snow,1\au}$ is independent of $\alpha_{\rm D}$ for $\amax \lesssim 1\um$,   $t_{\rm snow,\,1\au}$ decreases with $\alpha_{\rm D}$ for $\amax \gtrsim 1\um$.
In the case of $\alpha_{\rm D}=10^{-5}$, the snow line passes $1\au$ within 1.5 Myr even for $\amax=10$--$100\um$, for which accretion heating is the most efficient.

The dependence on $\alpha_{\rm D}$ reflects the effect of dust settling on the disk opacity $\kappa_{\rm R}${, which depends more on large grains than the heating layer height does}. 
As discussed in Section~\ref{subsec:iteration}, the mean opacity is dominated by grains with $a \la \lambda_{\rm peak} \sim 10~\rm \mu m$. With $\alpha_{\rm D}=10^{-5}$, grains larger than 10$\um$ settle substantially (see Section~\ref{subsec:dust}), causing the decrease of the opacity at $z=z_{\rm heat}$ for $\amax \ga 10~\um$. This is illustrated in the lower panel of Figure~\ref{fig:alpha}, which compares $\kappa_{\rm R}$ at $z=z_{\rm heat}$, $r=1\au$, and $t=0.6\Myr$ as a function of $\amax$ for $\alpha_{\rm D} = 10^{-3}$, $10^{-4}$, and $10^{-5}$. 
The decrease of the opacity at the heating layer results in the decrease of the accretion heating efficiency and hence the decrease of $t_{\rm snow, 1\au}$.
The heating layer height $z_{\rm heat}$ hardly depends on $\alpha_{\rm D}$ because the disk ionization structure is primarily controlled by the smallest, $0.1\um$-sized grains, which do not settle even with $\alpha_{\rm D}=10^{-5}$.

\subsubsection{Dependence on $f_{\rm dg}$}\label{sssec:fdg}

\begin{figure}
    \centering
    \includegraphics[width=\hsize, bb=0 0 281 227]{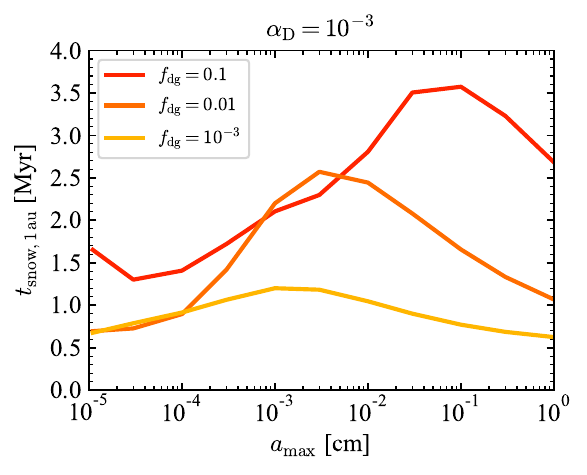}
    \includegraphics[width=\hsize, bb=0 0 272 227]{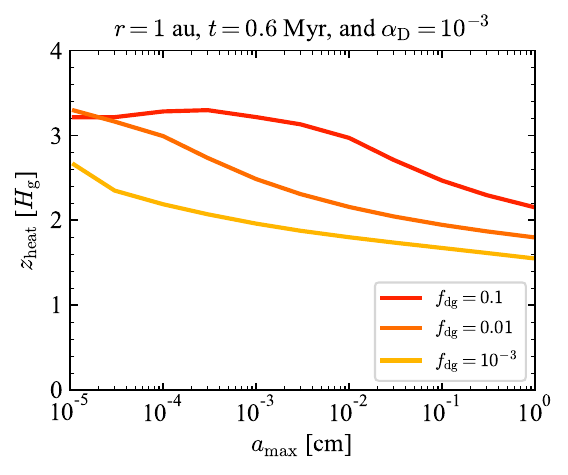}
    \caption{
    Snow line's arrival time at $1\au$ (upper panel) and heating layer height $z_{\rm heat}$ at $r=1\au$ and $t=0.6\Myr$ (lower panel) as a function of $\amax$ for different values of the dust-to-gas mass ratio $f_{\rm dg}$ with $\alpha_{\rm D}=10^{-3}$.
    }
    \label{fig:fdg}
\end{figure}
So far, we have fixed the dust-to-gas surface density ratio $f_{\rm dg}$ to the interstellar dust-to-gas mass ratio of 0.01 {\citep{Bohlin+1978}}. As mentioned in Section~\ref{subsec:dust}, $f_{\rm dg}$ in the inner $\sim 1~\rm au$ region can differ greatly from the interstellar value because of the grains' radial drift.  
The upper panel of \figref{fig:fdg} shows how $t_{\rm snow,\,1\au}$ depends on $f_{\rm dg}$ in the particular case of $\alpha_{\rm D}=10^{-3}$.
Overall, we find that accretion heating becomes inefficient as $f_{\rm dg}$ decreases. This is simply because the disk opacity $\kappa_{\rm R}$ decreases with $f_{\rm dg}$ (see \eqref{eq:kappa}). The heating layer height also decreases (see the lower panel of Figure~\ref{fig:fdg}), but inspection shows that the decrease in $\kappa_{\rm R}$ dominates over the increase in the column density above the heating layer.

It is interesting to note that unlike in the cases of $f_{\rm dg} \leq 0.01$, $t_{\rm snow,\,1\au}$ for $f_{\rm dg}=0.1$ is maximized at $\amax \sim 0.1\cm$, not at $\amax \sim 100~\rm \mu m$.
The increase of $t_{\rm snow,\,1\au}$ at $100\,{\rm \mu m} \lesssim \amax \lesssim 0.1$ cm is due to the relatively rapid decrease of $z_{\rm heat}$ from $3H_{\rm g}$ to $2H_{\rm g}$ (see the lower panel of Figure~\ref{fig:fdg}), which increases $\tau_{\rm heat}$ even though $\kappa_{\rm R}$ decreases. Around these heights, the ionization rate $\zeta$ has a shallow vertical profile owing to the contribution of scattered X-rays \citep{I&G1999,B&G2009} (see Figure~\ref{fig:zeta}).
This yields a shallow vertical Am profile around $z \sim 2$--$3H_{\rm g}$ (this can also be seen in Figure~\ref{fig:Am}). When ${\rm Am}$ crosses 0.3 around these heights, the value of $z_{\rm heat}$ (which is defined as the height where Am $= 0.3$) is sensitive to a change in Am with changing $\amax$. This explains why $z_{\rm heat}$ decreases steeply from  $3H_{\rm g}$ to $2H_{\rm g}$ as $a_{\rm max}$ increases from $100~\rm \mu m$ to 0.1 cm. 
Beyond $\amax \sim 0.1$ cm, $z_{\rm heat}$ stalls around $2H_{\rm g}$ because of the steep drop of $\zeta$ below this height, and consequently $t_{\rm snow,\,1\au}$ decreases as in the fiducial case for $\amax \ga 100~\rm \mu m$ (Section~\ref{sssec:amax}).

\subsubsection{Dependence on $f_{\rm heat}$ and $f_{\rm depth}$}\label{sssec:fdepth}

\begin{figure}
    \centering
    \includegraphics[width=\hsize, bb=0 0 281 227]{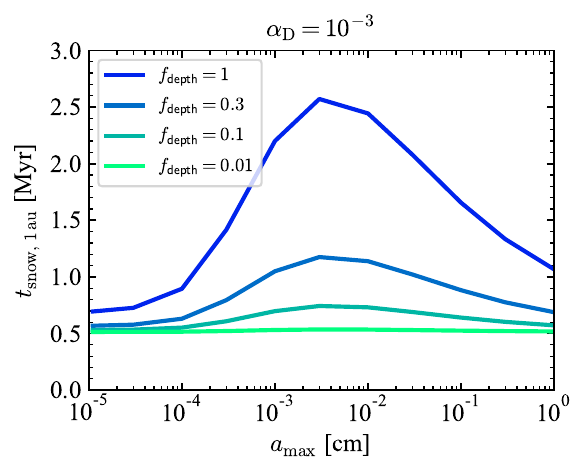}
    \caption{
    Snow line's arrival time at $1\au$ as a function of $\amax$ for different values of $f_{\rm depth}$ with $\alpha_{\rm D}=10^{-3}$. 
    }
    \label{fig:depth}
\end{figure}
Finally, we explore how the uncertainties in $f_{\rm heat}$ and $f_{\rm depth}$  affect our results.
As long as $\tau_{\rm heat} \ga 1$, these parameters affect $T_{\rm acc}$ equally, i.e., ($T_{\rm acc}\propto (f_{\rm heat}f_{\rm depth})^{1/4}$ (see  Equations (\ref{eq:acc}) and (\ref{eq:tau})), so it is sufficient to vary one of the two while fixing the other.  
\figref{fig:depth} shows $t_{\rm snow,\,1\au}$ as a function of $\amax$ for different values of $f_{\rm depth}$ with $\alpha_{\rm D}=10^{-3}$ and $f_{\rm heat} = 1$. 
As is evident from \eqref{eq:acc}, a lower $f_{\rm depth}$ leads to less efficient accretion heating, resulting in earlier snow line migration. In the extreme case of $f_{\rm depth} = 0.01$, accretion heating has no effect on $t_{\rm snow,\,1\au}$ for all values of $a_{\rm max}$. 
The same argument could be made for the case where $f_{\rm heat} < 1$ and $f_{\rm depth}=1$.

Therefore, we conclude that $f_{\rm heat}f_{\rm depth} { \approx} 1$ is needed for substantial accretion heating. This implies that the disk thermal structure and snow line migration critically depend on the Hall effect, which controls both $f_{\rm heat}$ and $f_{\rm depth}$. We discuss this point in more detail in Section~\ref{subsec:Hall}.


\section{DISCUSSION}\label{sec:dis}

\subsection{Effect of Grain Porosity}\label{subsec:porosity}
We have approximated dust grains with compact spheres of zero porosity. In reality, it is more likely that large dust grains forming in disks are aggregates of (sub)micron-sized grains. Aggregates generally have pores \citep[e.g.,][]{Blum08,Guttler10,Zsom11}, and can become significantly porous if the grains stick together through low-velocity collisions \citep[e.g.,][]{Dominik97,OkuzumiTanaka09,Kataoka13}. 

However, it can be shown that the results of this study are applicable, at least approximately, to porous aggregates if we replace $\amax$ by $f_{\rm fill}\amax$, where $f_{\rm fill}$ is the aggregates' filling factor. Here, we show this for a simplified case where all aggregates have the same radius $a$ and same internal density $\rho_{\rm int} \propto f_{\rm fill}$. 
\citet{Kataoka2014} show that an aggregate's absorption opacity is approximately determined by its mass-to-area ratio $m_{\rm d}/\pi a^2 \propto f_{\rm fill} a$. This is also the case for the aggregate's Stokes number ${\rm St} \propto \rho_{\rm int}a \propto f_{\rm fill}a$ (see \eqref{eq:St}). 
The disk ionization fraction is mainly controlled by the total cross section of grains per unit volume, ${\cal A}_{\rm tot} = \rho_{\rm d} \pi a^2/m_{\rm d} \propto \rho_{\rm d}/(f_{\rm fill}a)$ (see \eqref{eq:Atot}), which is also determined by $f_{\rm fill}a$ because $ \rho_{\rm d} \propto 1/H_{\rm d}({\rm St})$ for given $\Sigma_{\rm d}$. 
Therefore, the optical depth to the heating layer, which is determined by the opacity and heating layer height, is a function of the product $f_{\rm fill}a$.

\subsection{Can Vertical Dust Diffusion of $\alpha_{\rm D}> 10^{-5}$ Occur Around the Snow Line?}\label{subsec:alpha}
We have shown in Section~\ref{sssec:alpha} that a moderate level of vertical dust diffusion is necessary to make MHD accretion heating efficient. 
This requirement arises because the midplane temperature is determined by the optical depth above the heating layer, which lies at $z \sim 2$--$3H_{\rm g}$. Weak turbulence of $\alpha_{\rm D} > 10^{-5}$ is needed to lift opacity-dominating grains to such a high altitude. 
It is important to discuss whether there are any mechanisms that can produce such moderate turbulence in the inner disk region. 
Because MHD instabilities are likely suppressed around the snow line, we must look for non-MHD mechanisms. 

A potential candidate is the convective overstability, which is a hydrodynamical instability in a disk region with a negative radial entropy gradient and a thermal relaxation time comparable to the local Keplerian time \citep{K&H2014,Lyra2014}. The requirement for the thermal relaxation can be fulfilled around the snow line  \citep{Malygin2017,P&K2019}. A negative entropy gradient can also occur if accretion heating is effective.  
Simulations by \citet{Raettig2021} show that the convective overstability produces turbulence with $\alpha_{\rm D} \sim 10^{-5}$--$10^{-4}$ (see their Table 2, where $\delta$ corresponds to our $\alpha_{\rm D}$).

Another candidate is the vertical shear instability, which is a hydrodynamical instability caused by a vertical gradient of the gas rotation velocity with a cooling time scale much shorter than the orbital time scale \citep{UrpinBrandenburg98,Nelson+13,LinYoudin15}.
This instability causes particularly strong vertical diffusion amounting to $\alpha_D \sim 5\times 10^{-3}$ \citep{Flock+20}. Linear theory suggests that the vertical shear instability is most effective at $r > 10~\rm au$ \citep{Malygin2017,L&U2019,P&K2019}, but recent simulations show that the vertical shear instability in the upper layers can still cause vertical diffusion near the midplane around the snow line \citep{PfeilKlahr2021}.
Importantly, because the gas motion produced by this instability is strongly vertically elongated, it is much less efficient at transporting angular momentum radially \citep[$\alpha \sim 0.1\alpha_{D}$ ;][]{Flock+20}. Therefore, its direct contribution to accretion heating may be limited. Our finding suggests that even if this is the case, the vertical shear instability can still contribute to disk accretion heating indirectly, by enhancing the optical thickness above the heating layer produced by MHD wind-driven accretion as considered in this study.

\subsection{Importance of the Hall Effect}\label{subsec:Hall}
The results presented in Section~\ref{sssec:fdepth} imply a potentially important role the Hall effect might play in shaping the radial compositional gradient of planetary systems. 
As demonstrated in Figure~\ref{fig:depth}, the efficiency of MHD accretion heating depends significantly on $f_{\rm depth}$, with $f_{\rm depth} = 0.1$--$0.01$ being insufficient for the accretion heating to affect $t_{\rm snow, 1\au}$. This means that early snow line migration is inevitable in disks with anti-aligned vertical magnetic fields and rotation axis. Considering pebble accretion, we expect that planetary embryos that form around 1 au in disks with anti-aligned vertical magnetic fields would be more water-rich than those at the same orbit in disks with aligned vertical magnetic fields.

Protoplanetary disks' vertical magnetic fluxes are thought to be inherited from their parent molecular clouds. \citet{Tsukamoto2015} already predicted that during the collapse of prestellar cores, the Hall effect would result in the bimodal size distribution of the initial disks, assuming that the relative orientation between each core's magnetic flux and rotation axis is randomly determined. The results presented in this study suggest that the Hall term could also lead to bimodal evolution of the disk temperature distribution and snow line. As discussed above, this could in turn give rise to the bimodal radial distribution of exoplanets' water content.
The bimodal nature of the Hall term may also affect the fate of planets' migration and pebble accretion in disks, both of which depend on whether accretion heating is efficient or not  \citep{Bitsch19}. Future simulations of planet growth and migration combined with our MHD disk model will clarify the roles of the Hall term in planet formation.

\subsection{Dust Growth and Planetesimal Formation}\label{subsec:planetesimal}

In this paper, we have treated the maximum grain size as a free parameter.
Physically, the maximum grain size is determined by processes (often referred to as ``barriers'') limiting local dust growth, including fragmentation and radial inward drift.
If the sticking efficiency of the colliding grains is low, the maximum grain size is determined by collisional fragmentation \citep{Birnstiel09}.
Below, we estimate the maximum grain size limited by the fragmentation barrier and discuss how it constrains the snow line evolution as well as planetesimal formation.

Assuming that Epstein's law applies, the fragmentation-limited grain size can be expressed as \citep{Birnstiel09,Birnstiel12}
\begin{equation}
    a_{\rm frag}=\frac{2}{3\pi}\frac{\Sigma_{\rm g}}{\rho_{\rm int}\alpha_{\rm turb}}\frac{v_{\rm frag}^2}{c_{\rm s}^2},\label{eq:a_frag}
\end{equation}
where $\alpha_{\rm turb}$ is a dimensionless coefficient characterizing turbulence strength and $v_{\rm frag}$ is the fragmentation threshold velocity above which the colliding grains fragment rather than stick.
The parameter $\alpha_{\rm turb}$ is defined such that the velocity dispersion of the largest turbulent eddies is $\sqrt{\alpha_{\rm turb}}c_{\rm s}$ \citep{OrmelCuzzi07}, and is expected to be comparable to the dimensionless diffusion coefficient $\alpha_{\rm D}$ used in this paper \citep[e.g.,][]{OkuzumiHirose11}.  
Using $c_{\rm s}=\sqrt{k_{\rm B}T/\mu m_{\rm p}}$, 
Equation~(\ref{eq:a_frag}) can be rewritten as 
\begin{eqnarray}
    a_{\rm frag} \approx  0.8 {\rm \, mm} && \pf{\Sigma_{\rm g}}{400\,{\rm g\,cm^{-2}}}\pf{\rho_{\rm int}}{1.46 {\rm \, g\,cm^{-3}}}^{-1}\nonumber\\
    &&\pf{\alpha_{\rm turb}}{10^{-3}}^{-1}\pf{v_{\rm frag}}{1 {\rm \,m\,s^{-1}}}^2\pf{T}{200\,{\rm K}}^{-1}.\label{eq:frag}
\end{eqnarray}
The threshold sticking velocity for silicates is highly uncertain and can range between 1 $\rm m~s^{-1}$ \citep{Blum08} to 50 $\rm m~s^{-1}$ \citep{Kimura15} depending on the size and surface adhesion of the unit grains. 
If we take $v_{\rm frag} = 1~{\rm m~s^{-1}}$, then Equation~(\ref{eq:frag}) implies that the maximum grain size in the vicinity of the snow line is $\sim$ mm.
According to Figure~\ref{fig:time_snow}, this maximum size is larger than the sweet spot (10--100~$\micron$) where accretion heating becomes most efficient, but can still delay the  snow line's arrival to 1 au by up to a few Myr compared to the case where dust grains remain to be $0.1\,{\rm \mu m}$ in size.
Accretion heating would be less efficient for higher values of $v_{\rm frag}$.
In any case, accounting for dust growth is essential for elucidating how fast the snow line migrates in magnetically accreting disks. This will be the subject of our future work.

Although the suppression of dust growth at millimeter sizes is beneficial for the delay of snow line migration, it could also suppress planetesimal formation. Here, we briefly discuss whether planetesimal formation is possible in the scenario with $a_{\rm max} \sim 1~\rm m~s^{-1}$.  
To date, the streaming instability \citep{YoudinGoodman05} is one of the leading mechanisms to concentrate dust particles and produce planetesimals. 
The streaming instability is caused by the relative drift of dust and gas and triggers strong clumping of dust when the dust-to-gas mass ratio is high \citep[e.g.,][]{JohansenYoudin07,Carrera15,Yang17}.
Strong clumping takes place most vigorously when the grains are centimeter to meter-sized. 
However, \citet{Yang17} and \cite{LiYoudin21} show that the streaming instability can even concentrate relatively small, millimeter-sized grains if the dust-to-gas density ratio at the midplane is above a few (see Figure 4 of \citealt{LiYoudin21}). 
Their simulations treat an idealized situation where no externally driven turbulence like the one driven by the vertical shear instability is present.
Nevertheless, the possibility remains that the streaming instability can lead to planetesimal formation even in the presence of turbulence if there exists a mechanism that leads to $\rho_{\rm d}/\rho_{\rm g}\gtrsim 1$ at the midplane (although it would require a much higher dust-to-gas surface density ratio than in the absence of externally driven turbulence). This scenario is only speculative and needs to be verified by future simulations of the streaming instability.

In summary, to understand the snow line migration and rocky planetesimal formation in magnetically accretion protoplanetary disks in a consistent way, one needs to simulate the evolution of dust and disk temperature simultaneously. We plan to pursue this direction in future work.

\subsection{Can Planetary Embryos at 1 au Avoid Excessive Water Accretion?}\label{subsec:earth}

\cite{Mori2021} concluded that the snow line would migrate inside $1\au$ within 1~Myr if all dust grains contributing to the disk opacity and ionization were submicron-sized. 
In this study, we have found that the snow line's arrival at $1\au$ can be delayed, under some favorable conditions (see Sections~\ref{subsec:alpha} and \ref{subsec:Hall}), if the grains grow to $\amax=10$--$100\um$. The question then is whether planetary embryos lying at $\sim$ 1~au can remain dry in magnetically accreting disks. 

We cannot yet give a complete answer to this question because we still treat $\amax$ and the dust-to-gas mass ratio as a free parameter. According to the work by \cite{Sato2016}, who studied water delivery to Mars-sized rocky embryos via ice pebble accretion, the final water content of embryos depends on the icy pebbles' size and the radial extent of the protoplanetary disk, not only on the timing of the snow line passage. The disk size determines how quickly the ice pebbles' radial inward flux decays. Because of the high efficiency of pebble accretion, the Mars-sized embryos can only avoid excessive water accretion if the snow line passes them after the radial pebble flux has already decayed significantly. Moreover, the decay of the inward pebble flux is always accompanied by a decrease in the dust surface density, which would make accretion heating in the inner disk region less efficient (see Section~\ref{sssec:fdg}).
Simulations that self-consistently include dust evolution (growth, fragmentation, and radial drift), snow line evolution, and pebble accretion by planetary embryos are needed to answer the question of how dry rocky planets form in the inner region of magnetically accreting protoplanetary disks. Presenting such a simulation is beyond the scope of this study but should be done in future work.

\section{SUMMARY AND CONCLUSIONS}\label{sec:summary}

In this study, we have investigated the effects of the dust size and vertical distributions on the snow line migration in magnetically accreting protoplanetary disks.
In the disks, the gas motion is mainly laminar due to nonideal MHD effects, and thus the accretion heating would be caused by the Joule dissipation at the disk surface.
The disk temperature depends on the ionization fraction and disk opacity, 
which are determined by the dust size and spatial distributions. 
Therefore, the dust distributions can control the temperature evolution and snow line migration.
In addition, we have investigated the effect of the dust settling on the temperature structure.

Our key results are summarized  as follows.
\begin{enumerate}
    \item The disk temperature depends on the maximum grain size $\amax$ nonmonotonically and takes a maximum value at $\amax \sim 10$--$100\um$.
    The disk temperature is determined by the heating layer height and the opacity. 
    The heating layer height decreases monotonically with increasing $\amax$. 
    Thus, since the gas density profile is independent of $\amax$, 
    a larger $\amax$ leads to a larger column surface density at the heating layer. 
    The opacity around the heating layer is nearly constant at $\amax \lesssim 10\um$,
    but decreases with increasing $\amax$ at $\amax \gtrsim 10\um$. 
    Because of these effects, the disk temperature is also maximized at $\amax\sim 10$--$100\um$ (Section \ref{subsec:iteration}).

    \item The snow line's arrival at $1\au$ can be delayed up to $t \sim$ {2--3}$\Myr$ for $\amax \sim 10$--$100\um$,
    whereas the arrival time is $0.5\Myr$ for single $0.1\um$-sized dust grains.
    Even if $a_{\rm max} \sim$ mm, the snow line's arrival at 1 au is delayed to 1--2 Myr. Therefore, dust growth is an important factor that determines the location and evolution of the snow line (Section \ref{sssec:amax}).

    \item Dust sedimentation onto the midplane (i.e., a smaller $\alpha_{\rm D}$) results in the snow line's arrival at $1\au$ earlier.
    For $\alpha_{\rm D}=10^{-5}$, the snow line passes 1~au within 1.5~Myr even if $\amax$ lies in the range of 10--100$\um$. This dependence of $\alpha_{\rm D}$ comes from a smaller $\alpha_{\rm D}$ leading to a decrease in the opacity. 
    In particular, because $10\um$-sized grains dominate the opacity and settle substantially with $\alpha_{\rm D}=10^{-5}$, 
    the snow line passes $1\au$ earlier (Section \ref{sssec:alpha}). 
    
    \item 
    When the vertical diffusion is enhanced by some hydrodynamic instability ($\alpha_{\rm D} > 10^{-5}$ ), 
    the vertically diffused dust increases the opacity of the upper region,
    and therefore it makes the accretion heating efficient. 
    This suggests that turbulence can warm the disk even if its dissipation does not warm the disk directly.

    \item 
    When the direction of the magnetic field threading the disk is anti-aligned with the disk rotation axis,
    the heating layer height is located higher due to the Hall effect.
    In this case, the accretion heating in MHD disks is insufficient for all dust size distributions.
    The results in this study suggest that the Hall term could lead to bimodal evolution of the disk temperature distribution, the snow line migration, and in turn radial distribution of exoplanets' water content (Section \ref{subsec:Hall}).

\end{enumerate}

In this study, the dust size distribution is determined by the free parameters: the maximum grain size $\amax$ and the dust-to-gas surface density ratio $f_{\rm dg}$. 
However, the radial profile of the dust size distribution changes with time through radial drift, growth, and fragmentation of dust.
Thus, we should note that because $f_{\rm dg}$ would decrease at the inside of disks with time,
the snow line is expected to pass 1~au earlier than 2~Myr even at $a_{\rm max} \sim 10$--100$\um$.
Calculating $\amax$ and $f_{\rm dg}$ as a function of time $t$ considering those effects of dust,
we expect to understand more realistic snow line evolution in magnetically accreting disks.
Moreover, the water content of planetary embryos at 1~au is also determined by the ice pebble's size and the radial extent of a gas disk \citep[][see Section~\ref{subsec:earth} in this paper]{Sato2016}. 
Therefore, simulations self-consistently including the dust evolution, snow line evolution, and pebble accretion onto planetary embryos would answer the question of how dry rocky planets form, in the future.

\begin{acknowledgments}
The authors thank the anonymous referee for the many comments that greatly helped improve the manuscript. This work was supported by Japan Society for the Promotion of Science KAKENHI grant Nos.~JP18H05438, JP19K03926, JP19K03941, JP20H00182, JP20H00205, JP20H01948, and JP21J00086.
\end{acknowledgments}


\appendix

\setcounter{equation}{0}

\section{Expression of the ambipolar diffusivity}\label{asec:eta}
\setcounter{equation}{0}

The ambipolar diffusivity is given by \citep{Wardle2007}
\begin{equation}
    \eta_{\rm A}=\frac{c^2\sigma_{\rm P}}{4\pi (\sigma_{\rm P}^2+\sigma_{\rm H}^2)}-\eta_{\rm O},
\end{equation}
where $c$ is the speed of light, $\sigma_{\rm P}$ and $\sigma_{\rm H}$ are the Pedersen and Hall conductivities, and $\eta_{\rm O}$ is the ohmic diffusivity given by
\begin{equation}
    \eta_{\rm O}=\frac{c^2}{4\pi\sigma_{\rm O}},
\end{equation}
with $\sigma_{\rm O}$ being the ohmic conductivity.
The ohmic, Hall, and Pedersen conductivities are written as
\begin{equation}
    \sigma_{\rm O}=\frac{ec}{B}\sum_\alpha |Z_\alpha|n_\alpha\beta_\alpha, 
\end{equation}
\begin{equation}
    \sigma_{\rm H}=-\frac{ec}{B}\sum_\alpha \frac{Z_\alpha n_\alpha \beta_\alpha^2}{1+\beta_\alpha^2}=\frac{ec}{B}\sum_\alpha \frac{Z_\alpha n_\alpha}{1+\beta_\alpha^2},
\end{equation}
\begin{equation}
\sigma_{\rm P}=\frac{ec}{B}\sum_\alpha \frac{|Z_\alpha| n_\alpha \beta_\alpha}{1+\beta_\alpha^2},
\end{equation}
where $Z_\alpha$ and $n_\alpha$ are the charge number and number density of species $\alpha$, and
\begin{equation}
    \beta_{\alpha}=\frac{eB|Z_\alpha|\tau_\alpha}{m_\alpha c},
\end{equation}
is the Hall parameter with $m_\alpha$ and $\tau_\alpha$ being the mass and stopping time of species $\alpha$. Because the inertia of charged grains with $a \la 0.1~\rm \mu m$ is much smaller than those of ions and electrons, we neglect the contribution of charged grains to the conductivities.
The stopping time of ions and electrons are written as \citep{Sano2000}
\begin{equation}
    \tau_{\rm i}^{-1}=n_{\rm g}\langle\sigma v\rangle_{\rm i}\frac{m_{\rm g}}{m_{\rm i}},
\end{equation}
\begin{equation}
    \tau_{\rm e}^{-1}=n_{\rm g}\langle\sigma v\rangle_{\rm e},
\end{equation}
where $\langle\sigma v\rangle_{\rm i,(e)}$ is the rate coefficient for the collision between ions (electrons) and neutrals averaged over the distribution of their relative velocity.

\section{Expression of the Ionization Rate}\label{asec:zeta}

We consider disk ionization by cosmic rays \citep{U&N2009}, stellar X-rays \citep{I&G1999}, and radionuclides \citep{U&N2009}.
We write the total disk ionization rate $\zeta$ as
\begin{equation}
    \zeta = \zeta_{\rm CR}+\zeta_{\rm XR}+\zeta_{\rm RA},
\end{equation}
where $\zeta_{\rm CR}$, $\zeta_{\rm XR}$, and $\zeta_{\rm RA}$ are the contributions from cosmic rays, stellar X-rays, and radionuclides, respectively.

The cosmic ray ionization rate is given by \citep{U&N2009}
\begin{eqnarray}
    \zeta_{\rm CR}(r,z)&=&\frac{\zeta_{\rm CR,0}}{2}\left\{\exp\left(-\frac{\Sigma_{\rm g}^+(r,z)}{\Sigma_{\rm CR}}\right)\left[1+\left(\frac{\Sigma_{\rm g}^+(r,z)}{\Sigma_{\rm CR}}\right)^{3/4}\right]^{-4/3}\right. \nonumber \\
    &+&\left.\exp\left(-\frac{\Sigma_{\rm g}^-(r,z)}{\Sigma_{\rm CR}}\right)\left[1+\left(\frac{\Sigma_{\rm g}^-(r,z)}{\Sigma_{\rm CR}}\right)^{3/4}\right]^{-4/3}\right\},
\end{eqnarray}
where $\zeta_{\rm CR,0}=1.0\times 10^{-17}\,{\rm s^{-1}}$ is the ionization rate of $\rm H_2$ in the interstellar space, $\Sigma_{\rm CR}=96\,{\rm g\,cm^{-2}}$ is the mean attenuation length of cosmic rays, and 
\begin{equation}
	\Sigma^{+}_{\rm g}(r,z)=\int_z^\infty \rho_{\rm g}(r,z')\d z', 
\end{equation}  
\begin{equation}
	\Sigma^{-}_{\rm g}(r,z)=\Sigma_{\rm g}(r)-\Sigma^{+}_{\rm g}(r,z),
\end{equation}
are the vertical gas column densities measured from the upper and lower infinities, respectively.

The radionuclide ionization rate is taken to be 
$\zeta_{\rm RA} = 7 \times 10^{-19}\,{\rm s^{-1}}$, which corresponds to the ionization rate produced by $^{26}{\rm Al}$ with an abundance ratio of $^{26}{\rm Al}/^{27}{\rm Al}=5 \times 10^{-5}$ \citep{U&N2009}.

For the X-ray ionization rate, we use an analytic expression by \cite{B&G2009} with some modifications,
\begin{eqnarray}
    &&\frac{\zeta_{\rm XR}(r,z)}{L_{\rm XR,29}}\pf{r}{1 \au}^{2.2} = \nonumber \\
    &&\zeta_{\rm XR,1}\left\{\exp\left[-\pf{\Sigma_g^+(r,z)}{\Sigma_{\rm XR,1}}^\alpha\right]+\exp\left[-\pf{\Sigma_g^-(r,z)}{\Sigma_{\rm XR,1}}^\alpha\right]\right\} \nonumber\\
    &&+\zeta_{\rm XR,2}\left\{\exp\left[-\pf{\Sigma_g^+(r,z)}{\Sigma_{\rm XR,2}}^\beta\right]+\exp\left[-\pf{\Sigma_g^-(r,z)}{\Sigma_{\rm XR,2}}^\beta\right]\right\},
\end{eqnarray}
where $L_{\rm XR,29}\equiv L_{\rm XR}/(10^{29}\ergs)$, $\zeta_{\rm XR,1}=4.0\times10^{-12} \,{\rm s^{-1}}$, $\Sigma_{\rm XR,1}/(1.42m_{\rm p})=1.0\times10^{21}\,{\rm cm^{-2}}$, $\alpha=0.4$, $\zeta_{\rm XR,2}=1.0\times10^{-15}\,{\rm s^{-1}}$, $\Sigma_{\rm XR,2}/(1.42m_{\rm p})=1.0\times10^{24}\,{\rm cm^{-2}}$, and $\beta=0.7$. We have modified the parameter values such that the formula better reproduce  Figure 3 of \cite{I&G1999} for the 5 keV case.
In this study, we assume the stellar X-ray luminosity $L_{\rm XR}$ of $10^{30}\ergs$.

\section{Evolution of the midplane temperature}\label{asec:T_ev}
\begin{figure*}
    \centering
    \includegraphics[width=\hsize,clip, bb=0 0 857 284]{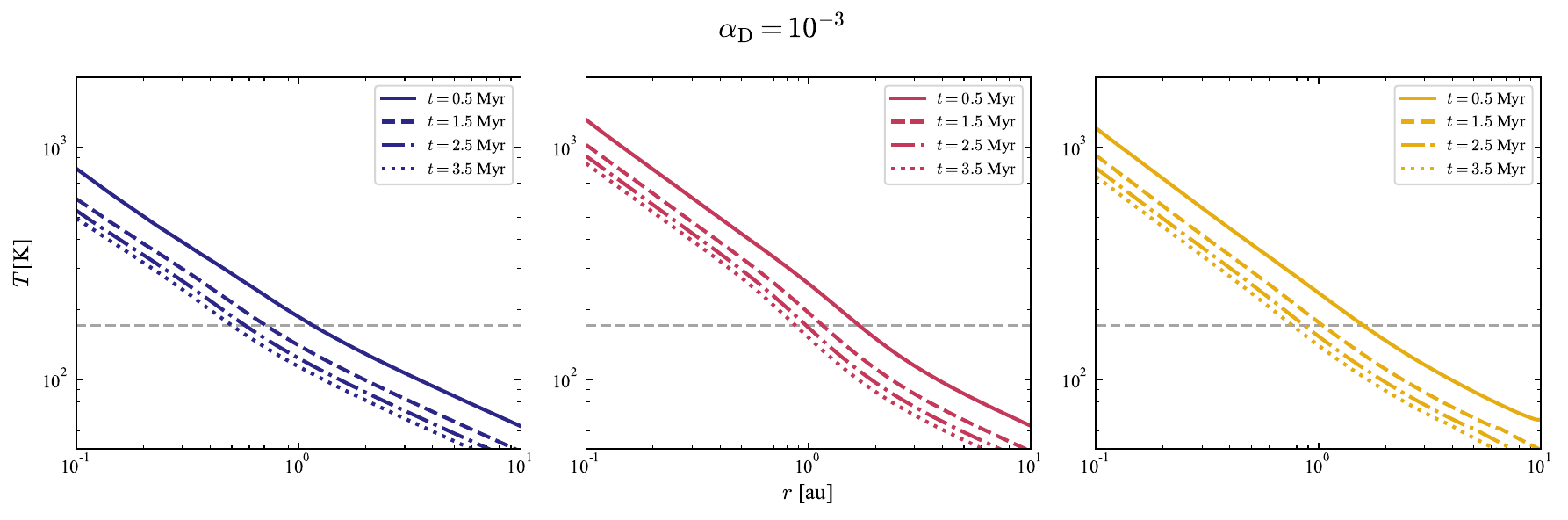}
     \caption{Radial profiles of the midplane temperature $T$ at different times for $\alpha_{\rm D}=10^{-3}$. The dashed horizontal line marks the water ice sublimation temperature $T=170\K$ assumed in this study. Left, middle, and right panels are for $\amax=0.11\,{\rm \um}$, $\amax=10\,{\rm \um}$, and $\amax=0.1$ cm, respectively.}
    \label{fig:T_ev}
\end{figure*}
Figure~\ref{fig:T_ev} shows the radial profiles of the midplane temperature $T$ at different times for $\alpha_{\rm D}=10^{-3}$.

\bibliography{Kondo}

\end{document}